\documentclass[twocolumn]{aastex631}
\usepackage[normalem]{ulem}
\usepackage{amsmath}
\usepackage{graphicx}
\usepackage{CJK}
\usepackage{multirow}
\usepackage{rotating}
\usepackage{tikz}

\begin{document}
\begin{CJK*}{UTF8}{gbsn}

\title{Observational characteristics of circum-planetary-mass-object disks in the era of James Webb Space Telescope}

\author[0009-0005-0743-7031]{Xilei Sun (孙锡磊)}
\affiliation{School of Physics and Astronomy, Sun Yet-sen University, Zhuhai 519082, People's Republic of China}

\author[0000-0002-7575-3176]{Pinghui Huang (黄平辉)}
\affiliation{Department of Physics and Astronomy, University of Victoria, Victoria, BC V8P 5C2, Canada}

\author[0000-0001-9290-7846]{Ruobing Dong (董若冰)}
\affiliation{Department of Physics and Astronomy, University of Victoria, Victoria, BC V8P 5C2, Canada}

\author[0000-0002-9442-137X]{Shang-Fei Liu (刘尚飞)}
\affiliation{School of Physics and Astronomy, Sun Yet-sen University, Zhuhai 519082, People's Republic of China}
\affiliation{CSST Science Center for the Guangdong-Hong Kong-Macau Greater Bay Area, Sun Yat-sen University, Zhuhai 519082, China}
\email{liushangfei@mail.sysu.edu.cn}



\begin{abstract}

Recent observations have confirmed circumplanetary disks (CPDs) embedded in parental protoplanetary disks (PPDs). On the other hand, planetary-mass companions (PMCs) and planetary-mass objects (PMOs) are likely to harbor their own accretion disks. Unlike PPDs, CPDs and other disks around planet analogues are generally too compact to be spatially resolved by current instrumentation. In this study, we generate over 4,000 spectral energy distributions (SEDs) of circum-PMO-disks (CPMODs) with various host temperature and disk properties, which can be categorized into four prototypes, i.e., full, pre-transitional, transitional and evolved CPMODs. We propose a classification scheme based on their near-to-mid-infrared colors. Using those CPMOD models, we synthesize JWST (NIRCam and MIRI) photometry for F444W, F1000W and F2550W wide filters. We show F444W - F1000W and F444 - F2550W colors can be applied to distinguish different types of CPMODs, especially for those around hot hosts. Our results indicate that the ongoing and future JWST observations are promising to unveil structures and properties of CPMODs.

\end{abstract}



\section{Introduction} \label{sec:intro}

    Despite a long hypothesized supposition that a circumplanetary disk (CPD) is formed when a newly formed planet opens a gap in the protoplanetary disk (PPD), detection of millimeter and submillimeter emissions of CPDs embedded in parental circumstellar disks remains very challenging \citep{Andrews_2021}. Currently, only a handful attempts yields positive results, including compact continuum emission sources associated with two candidate protoplanets in PDS 70 system \citep{Isella_2019,Wang_2021,Benisty_2021} and a point source in molecular line emission embedded in the AS 209 disk \citep{Bae_2022}. However, candidate CPDs in these systems are not spatially resolve due to their small sizes even with the unparalleled Atacama Large Millimeter/submillimeter Array (ALMA).

    CPDs are believed to be the cradles of moons, such that the four Galilean satellites could form in a closely packed coplanner configuration around Jupiter \citep[see e.g.][and references therein]{2010AJ....140.1168W, Fung_2019, Chen_2020, 2023ASSL..468...41M}. 
    One of the most intriguing questions about CPDs is how satellites form.  
    \cite{Fung_2019} and \cite{2022ApJ...939L..23C} have explored the patterns of gas flow around disk-embedded planets with 3D hydrodynamic simulations. They find CPDs could be disky or spherical assuming different temperature profiles and satellite formation could be complicated by properties of the disk and the planet.
   
    Recent observations of planetary-mass companions (PMCs) suggest the presence of accretion disks around them, such as GSC 6214-210 b, DH Tau b \citep{Zhou_2014,Wolff_2017}, SR 12 c \citep{Santamar_a_Miranda_2017,Wu_2022}, Delorme 1 (AB) b \citep{Betti_2022,Ringqvist_2023}, TWA 27B \citep{Luhman_2023}, and GQ Lup B \citep{cugno2024midinfrared}. In addition, some planetary-mass objects (PMOs) in NGC 1333 have been identified as possibly having disks, evidenced by infrared emission exceeding estimated photospheric levels \citep{2023AJ....165..196S}. The formation of disks around free-floating planets (FFPs) is also a possibility \citep{Miret_Roig_2021}. 
    
    Properties of these planetary analogues (Table \ref{tab:ref}) are consistent with the hot start scenario \citep{Spiegel_2012}. In addition, most of them are accreting at a rate less than $10^{-7}\;M_\mathrm{J}\; \mathrm{yr}^{-1}$ suggesting host irradiation dominates accretion luminosity \citep{2015ApJ...799...16Z}. Although similar to CPDs in terms of configurations and sizes, disks around these substellar objects lack gas and dust feeding from parental circumstellar disks, making them potentially different systems. For instance, parental disk material falling onto the surface of a CPD at a supersonic speed induces shocks \citep{2012ApJ...747...47T}, which may significantly influence disk thermal and ionization structures. In this paper, we use the term circum-planetary-mass-object disks (CPMODs) to refer to all circum-substellar disks that are NOT embedded in a protoplanetary disk. In that sense, CPMODs may resemble more scaled-down versions of circumstellar disks than CPDs. Nonetheless, CPMODs still offer a valuable viewpoint to study grain growth and satellite formation around very low mass objects. 

     \begin{table*}[ht!]
        \centering
        \caption{Properties of PMOs and circum-PMO disks}
            \centering
                \begin{tabular}{c|c|c|c|c|c|c}
                    \hline
                    \hline
                    \multirow{2}{*}{\textbf{Source}} & \multirow{2}{*}{\textbf{Mass}$/M_\mathrm{J}$} & \multirow{2}{*}{\textbf{Radius}$/R_\mathrm{J}$} & \textbf{Temperature} & \textbf{Accretion rate} \footnote{Note that the estimation of accretion rate sensitively depends on models.} & \textbf{Disk total} &\multirow{2}{*}{\textbf{Reference}} \\
                    & & & $/\mathrm{K}$ & $/M_\mathrm{J} \;\mathrm{yr}^{-1}$ & \textbf{dust mass$/M_\mathrm{J}$} & \\
                    \hline
                    DH Tau b & $11 \pm 3$ & $2.7 \pm 0.8$ & $\sim2200$ & $3.2 \times 10^{-9}$ & $\lesssim 1.3\times10^{-3}$ & \citet{Zhou_2014}; \\
                    \cline{1-6} 
                    GSC 6214-210 b & $15 \pm 3$ & $1.8 \pm 0.5$ &$\sim2200$ & $1.6 \times 10^{-8} $ &$\lesssim 5\times10^{-4}$ &\citet{Wolff_2017} \\
                    \hline
                    GQ Lup B & $20 \pm 10$ & $3.6 \pm 0.1$ &$\sim2700$ & $3.2\times 10^{-7}$ & - &\citet{cugno2024midinfrared} \\
                    \hline
                    TWA 27B & $5.5 \pm 0.5$ & $1.3 \pm 0.2$ & $\sim1300$ & $\lesssim10^{-9}$ & $\lesssim 4\times10^{-5}$& \citet{Luhman_2023} \\
                    \hline
                    SR 12 c & $11 \pm 3$ & $1.6 \pm 0.7$ & $2600\pm 100$ & $10^{-8}$ &$\lesssim 1.7\times10^{-4}$ &\citet{Wu_2022} \\
                    \hline
                    PDS 70 c \footnote{PDS 70 c is a protoplanet harboring its own CPD. We include it here for reference.} & $<10$& - & - & $10^{-8}$ & $\lesssim 1\times10^{-4}$ &\citet{Benisty_2021}\\
                    \hline
                    \hline
                \end{tabular}
            \label{tab:ref}
    \end{table*}

    Due to the nature of compactness of CPDs and CPMODs, state-of-the-art observations merely reveal their spectral energy distributions (SEDs). In particular, the infrared observation range of Near-Infrared Camer (NIRCam) and Mid-infrared instrument (MIRI) equipped with the James Webb Space Telescope (JWST), spanning $0.6 \sim 28.5 \:\mathrm{\mu m}$ \citep{2023PASP..135d8003W}, provides a great opportunity to observe these structures around young sub-stellar objects \citep{2024arXiv240310057C}. \citet{cugno2024midinfrared} report significant grain growth in GQ Lup B disk. At the time of writing, a number of JWST cycle 3 proposals have been approved to explore CPDs and CPMODs. Thus it is of particular relevance to predict what could be observed with infrared instrumentation. Inspired by the SED classification of PPDs \citep{1992ApJ...393..278L,Hernandez_2007,2014prpl.conf..497E}, we take a similar approach to classify CPMODs based on their SEDs assuming a disky structure and evolution model for simplicity. This new classification scheme may advise the ongoing CPMOD observations and help to unveil different structures of CPMODs.
    
    The paper is organized as follows, Section \ref{sec:method} outlines our parametric models, hydrodynamic models, and radiative transfer simulations. Section \ref{sec:result} encompasses the presentation of numerical simulation results and accompanying discussions. The concluding remarks for this paper are provided in Section \ref{sec:conclusion}.

\section{Numerical methods} \label{sec:method}

    In this section, we establish a series of parametric CPMOD models and conduct radiative transfer simulations to obtain their SEDs. We further verify our parametric CPMOD models through two additional hydrodynamic simulations, and we show their radiative transfer simulations are in good agreement with those based on the parametric models in Section \ref{sec:hdresult}.
    
    \subsection{Model Parameters of a full CPMOD} \label{sec:rtmodel}
    
        To investigate a vast parameter space efficiently, we assume that a CPMOD consists of a central planet and a surrounding dusty disk, which can be described by a few key variables listed in the full disk part of Table \ref{tab:model}.
        
        \begin{table*}[ht!]
            \centering
            \caption{Parameters of Circum-PMO Disk Models}
                \begin{tabular}{l c r}
                    \hline
                    \hline
                    \textbf{Parameter} & \textbf{Symbol}& \textbf{Value} \\
                    \hline
                    \textbf{Full disk} & \\
                    \hline
                    \textbf{Planet} \\
                    Effective temperature&$T_{\rm eff}$ & [1000,\:2000]\:\rm{K}\\
                    Mass & $M_\mathrm{p}$ & $10\: M_\mathrm{J}$\\
                    Radius & $R_\mathrm{p}$ & $2\: R_\mathrm{J}$\\
                    \\
                    \textbf{Dust grain}\\
                    Grain size\footnote{This study focuses on near-to-mid-infrared wavelengths up to $\sim$30 $\mu m$, which are mostly sensitive to small grains up to 10 $\mu m$.} & $[a_\mathrm{min},a_\mathrm{max}]$ & [0.1,\:10]$\:\mathrm{\mu m}$\\
                    Size power index & $q$ & 3.5 \\
                    \\
                    \textbf{Dust disk}\\
                    Reference radius &$R_0$ & $10\:R_\mathrm{J}$\\
                    Dust surface density at $R_0$ & $\Sigma_\mathrm{d,0}$ & [0.1,\:6] $\: \mathrm{g \cdot cm^{-2}}$ \\
                    Density power index&$\gamma$ & -1.5,\:-1.0,\:-0.8\\
                    Aspect ratio at $R_0$ & $h_\mathrm{d,0}/R_0$ & 0.05, 0.08\\
                    Flaring index &$\beta$ & 1.15,\:1.3\\
                    Model dust mass\footnote{Mass of dust grains within the size range $[a_\mathrm{min},a_\mathrm{max}].$ that contributes to near-to-mid-infrared flux up to $\sim 30 \mu m$.} &$M_\mathrm{d}$ & $(2 \times 10^{-8},\:1 \times 10^{-5})\: M_\mathrm{J}$\\
                    Total disk dust mass\footnote{Inferred total dust mass in disk with a broader size range $\left[ 0.1 \;\mu m , 1 \;\mathrm{mm} \right]$ that follows the same power law distribution $n(a)\propto a^{-q}$.} &$M_\mathrm{d, disk}$ & $(\sim 2 \times 10^{-7},\:\sim 1 \times 10^{-4})\: M_\mathrm{J}$\\                    \\
                    \textbf{Mesh}\\
                    Radial grid range& $R$ & $\left[5,\:500\right]\:R_\mathrm{J}$\\
                    Azimuthal grid range& $\phi$ & $\left[0,\:2\pi\right)$\\
                    Polar grid range&$\theta$ & $\left[0,\:\frac{\pi}{6}\right]$\\
                    Mesh resolution & $N_R \times N_\phi \times N_\theta $ &$100\:\times\:1\:\times\:32$\\
                    \\
                    \hline
                    \textbf{Other disks} &  \\
                    \hline
                    \textbf{Pre-transitional disk} \\
                    Gap inner edge &$R_\mathrm{in}$ & $10\:R_\mathrm{J}$\\
                    Gap outer edge &$R_\mathrm{out}$ & $50,\:100\:R_\mathrm{J}$\\
                    Gap depletion factor & $\Sigma_{\mathrm{gap}}/\Sigma_\mathrm{d}$ & $10^{-4},10^{-3}$\\
                    \\
                    \textbf{Transitional disk}\\
                    Cavity outer edge &$R_\mathrm{cavity}$ & $50,\:100,\:150,\:200\:R_\mathrm{J}$\\
                    Cavity depletion factor & $\Sigma_{\mathrm{cavity}}/\Sigma_\mathrm{d}$ & $10^{-4},10^{-3}$\\
                    \\
                    \textbf{Evolved disk}\\
                    Overall depletion factor& $\Sigma_\mathrm{E}/\Sigma_\mathrm{d}$ & $10^{-4},10^{-3}$\\
                    \hline
                    \hline
                \end{tabular}
            \label{tab:model}
        \end{table*}

        \begin{figure*}[t!]
            \centering
            \includegraphics[width=0.95\textwidth]{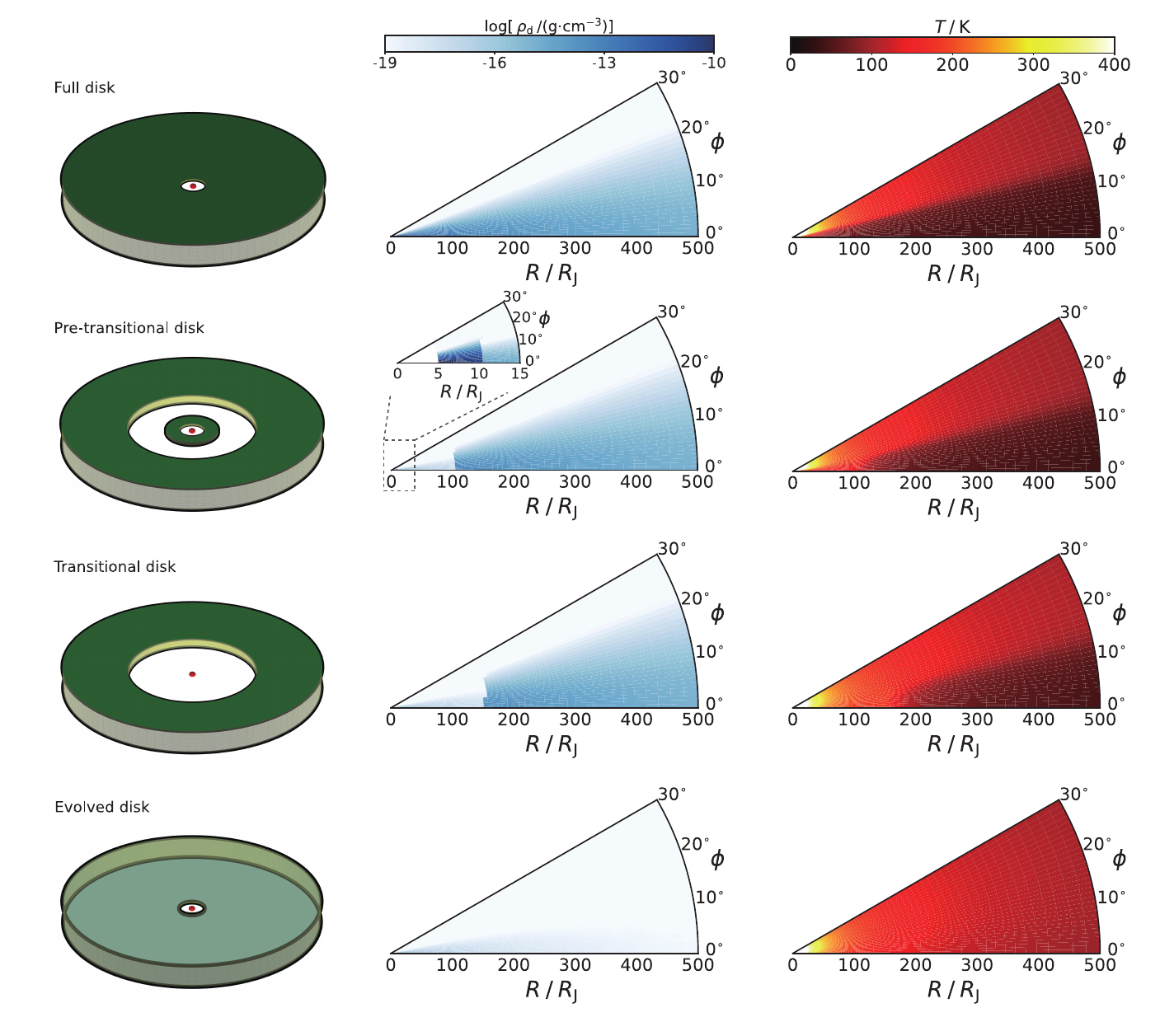}
            \caption{Model cartoons, density distributions and temperature distributions of four CPD prototypes. Each row represents a CPD parametric model. 
            \textbf{\emph{Left column:}} the 3D dust disk parametric model diagrams, where the green side represents the surface of the disk, the light yellow and grey side represents the side of the disk, and the red dot represents the planet in the disk center. The light green color in the bottom diagram indicates an optically thin disk.
            \textbf{\emph{Middle column:}} the dust density distribution in the r-$\theta$ plane. The central object is at $r=0$, the r direction ranges from 0 to 500 $R_\mathrm{J}$, and the $\theta$ direction is from 0 to 30 degree. The whole model is symmetrical about the $\theta = 0$ plane, so -30 to 0 degrees are not shown. 
            \textbf{\emph{Right column:}} the temperature distribution of the disk in the r-$\theta$ plane calculated by \texttt{RADMC-3D} using the thermal Monte-Carlo method (with $5\times10^8$ photons). The temperature of the central object assumed at 2,000 K.
            }
            \label{fig:model}
        \end{figure*}
        
        Guided by previous observations (Table \ref{tab:ref}), such as DH Tau b and its associated disk \citep{Wolff_2017}, we adopt a planet mass of $M_{\mathrm{p}}=10M_\mathrm{J}$, a planet radius of $R_{\mathrm{p}}=2R_\mathrm{J}$, and set the planet's temperature $T_{\mathrm{eff}}$ within the range [1000, 2000] K, where $M_\mathrm{J}$ and $R_\mathrm{J}$ denote the mass and radius of Jupiter, respectively.
            
        Because we are concerned with the dust continuum emission at near-to-mid-infrared wavelengths up to 30 $\mu m$ in this work, we simply need to study the distribution of dust grains up to 10 $\mu$m that dominate the dust opacity at wavelengths of interest. The prescription of dust grain size distribution will be discussed in Section \ref{sec:radiative}. 
        
        For a full disk (Figure \ref{fig:model}), the dust surface density profile is modeled by the formula
        \begin{equation}\label{eq:1}
            \Sigma_\mathrm{d} (R) = \Sigma_\mathrm{d,0} \cdot \left( \frac{R}{R_0} \right) ^ \gamma,
        \end{equation}
        where $R$ represents the radius, $\Sigma_\mathrm{d,0}$ represents the dust surface density at $R=R_0$, and $\gamma$ is the density power-law index. Simultaneously, in the vertical direction, the density distribution is assumed to follow the Gaussian vertical profile
        \begin{equation}\label{eq:2}
            \rho_\mathrm{d} (R, z) = \frac{\Sigma_\mathrm{d}(R)}{\sqrt{2\pi},h_\mathrm{d}(R)}\:\mathrm{exp}\left[-\frac{1}{2} \cdot \left(\frac{z}{h_\mathrm{d}(R)}\right)^2\right],
        \end{equation}
        where $R$ and $z$ represent the radius and height in cylindrical coordinates, respectively. $h_\mathrm{d}$ represents the dust scale height. The dust scale height $h_\mathrm{d}$  follows
        \begin{equation}\label{eq:3}
            h_\mathrm{d}(R)=h_\mathrm{d,0}\cdot \left( \frac{R}{R_0} \right) ^ \beta,
        \end{equation}
        where $h_\mathrm{d,0}$ is the dust scale height at distance $R_0$, and $\beta$ is the flaring index.

    \subsection{Prototypes of CPMODs}
        
        In this subsection, we introduce another three variations of CPMOD models along with the full disk model presented in the previous subsection. These four prototypes represent typical evolutionary stages among CPMODs, similar to the classification approach for PPDs \citep{2014prpl.conf..497E}, i.e., full disks, pre-transitional disks, transitional disks, and evolved disks (Figure \ref{fig:model}). We briefly summarize the characteristics of each prototype here.
       \begin{itemize}
            \item 
            A \textbf{full disk} has a continuous and smooth distribution of dust.
            \item 
            A \textbf{pre-transitional disk} is similar to a full disk except for that it has a wide gap in the middle, where dust is severely depleted.
            \item 
            A \textbf{transitional disk} has a effectively truncated inner cavity, so the disk inner edge is further away from the planet.
            \item 
            An \textbf{evolved disk} is like a full disk but with a much lower density throughout. 
        \end{itemize}

        As evidently shown in Figure \ref{fig:model}, the density distributions of four prototypes of CPMODs exhibit dramatic difference, which would be reflected in their SEDs. In practice, their dust surface density profiles are assumed to follow 
        \begin{equation}
            \Sigma_\mathrm{d,F} (R) = \Sigma_\mathrm{d,0} \cdot \left( \frac{R}{R_0} \right) ^ \gamma,
        \end{equation}
        \begin{equation}
            \Sigma_\mathrm{d,P} (R) = \begin{cases}\Sigma_\mathrm{d,F} (R) \cdot \left( \frac{\Sigma_\mathrm{gap}}{\Sigma_\mathrm{d}} \right),\quad & R_{\mathrm{in}} \le R \le R_{\mathrm{out}}\\ \Sigma_\mathrm{d,F} (R), \quad & R<R_{\mathrm{in}}\:\&\: R > R_{\mathrm{out}}
            \end{cases},
        \end{equation}
        \begin{equation}
            \Sigma_\mathrm{d,T} (R) = \begin{cases}\Sigma_\mathrm{d,F} (R) \cdot \left( \frac{\Sigma_\mathrm{cavity}}{\Sigma_\mathrm{d}} \right),\quad & R \le R_{\mathrm{cavity}}\\ \Sigma_\mathrm{d,F} (R), \quad & R > R_{\mathrm{cavity}} 
            \end{cases},
        \end{equation}
        \begin{equation}
            \Sigma_\mathrm{d,E} (R) = \Sigma_\mathrm{d,F} (R) \cdot \left( \frac{\Sigma_\mathrm{E}}{\Sigma_\mathrm{d}} \right),
        \end{equation}
        where the subscript F, P, T and E indicate each prototype, respectively. Note for the full disk prototype, the surface density profile follows Eq. \ref{eq:1}. The dust vertical density distribution Eq. \ref{eq:2} and dust scale height Eq. \ref{eq:3} still hold for all prototypes. Dust depletion in the latter three types of disks is defined by a depletion factor, i.e. $\Sigma_{\mathrm{gap}}/\Sigma_\mathrm{d}$ in gap, $\Sigma_{\mathrm{cavity}}/\Sigma_\mathrm{d}$ in cavity, and $\Sigma_\mathrm{E}/\Sigma_\mathrm{d}$ in whole disk, respectively. In this study, the depletion factor are set to be $10^{-4}$ and $10^{-3}$, representing a severe depletion and a moderate depletion \citep{Kanagawa_2015, Walker_2004}. The newly introduced parameters and their values are shown in the bottom part of Table \ref{tab:model}. In total, 4,620 models with different combinations of parameters are established in this work.

    \subsection{Radiative Transfer Simulations} \label{sec:radiative}
        
        We utilized the \texttt{RADMC-3D} code \citep{2012ascl.soft02015D} to conduct radiative transfer simulations for all CPMOD models. Planets are treated as blackbody radiations at different equilibrium temperatures for simplicity. \footnote{Note that the radiation of a 10 $M_\mathrm{J}$ planet formed under different conditions could vary and deviate from a pure blackbody to some extent \citep[see, e.g. Figure 6 from][]{Spiegel_2012}.} With the obtained temperature distributions of various disks, one can derive their SEDs accordingly.

        We model the dust grain size $a$ ranging from 0.1 $\mathrm{\mu m}$ to 10 $\mathrm{\mu m}$ following a steady state power-law size distribution $n(a) \propto a^{-q}$, where the power index $q=3.5$. Thus, the model dust mass in the full disk case is limited between $2 \times 10^{-8}$ and $1 \times 10^{-5} \: M_\mathrm{J}$, as shown in Table \ref{tab:model}. We adopt the DSHARP dust grain composition \citep{Birnstiel_2018}. And the dust opacities for absorption and scattering are computed with the \texttt{dsharp\_opac} routine based on Mie theory. The averaged dust opacity curves are shown in Appendix \ref{appendix:Dustopac}. 
        
        In reality, size of dust grains has a broader distribution. The choice of $[a_\mathrm{min},a_\mathrm{max}]$ in radiative transfer modeling is because we are interested in disk SED at near-to-mid-infrared wavelengths up to $\sim 30\;\mathrm{\mu m}$, such emission is mostly sensitive to the distribution of small dust up to 10 $\mu m$. Assuming the same power-law size distribution, the translated total dust mass in disk with a size ranging from 0.1 $\mathrm{\mu m}$ to 1 $\mathrm{mm}$ is about one-order-of-magnitude larger than the model dust mass, $M_\mathrm{d, disk}\sim 10 M_\mathrm{d}$ between $\sim 2 \times 10^{-7}$ and $\sim 1 \times 10^{-4} \: M_\mathrm{J}$ (Table \ref{tab:model}). This is comparable to the dust mass derived from observations (Table \ref{tab:ref}).

        The dust disk is simulated with a resolution of $100\times1\times32$ mesh grids along the radial, azimuthal and polar direction. The azimuthal direction is averaged out to optimize computational efficiency, as any potential asymmetric structure is not our primary focus given the low resolution of CPD observation. In the radial direction, the representation spans from 5 to 500 $R_\mathrm{J}$. For the polar direction, the model is set to be symmetric about the radial-azimuthal ($r-\theta$) plane to optimize computational efficiency, with the actual representation ranging from -30 to 30 degrees.
        
        We use $5\times10^5$ photons for ray tracing in each simulation and apply the thermal Monte Carlo method to derive the temperature distribution. The temperature results of four CPMOD prototypes are depicted in the right column of Figure \ref{fig:model}, with comprehensive analysis provided in Section \ref{sec:result}.

        Subsequently, \texttt{RADMC-3D} was employed to compute the SEDs for various models based on the temperature outcomes. The disk is face on (inclination angle $i=0^{\circ}$) and the distance to observer is set to 100 $\mathrm{pc}$. We present the SED results in Section \ref{sec:result}.

    \subsection{Hydrodynamic Models} \label{sec:hdmodel}
    
        We also constructed two CPMOD models using 2D hydrodynamic simulations with the \texttt{Athena++} code \citep{2020ApJS..249....4S} in cylindrical coordinates. This is primarily to investigate the gap formation mechanism in the pre-transitional disk prototype and to validate the rationality of our parametric models. The dust surface density including all grain sizes can be derived from the gas surface density assuming a constant dust-to-gas ratio of 1:100. As we have explained in the previous subsection, only small dust grains ranging from 0.1 $\mathrm{\mu m}$ to 10 $\mathrm{\mu m}$, roughly one tenth of the total dust mass, contribute to the near-to-mid-infrared wavelengths of interest. As a result, we obtain the dust surface density of small dust grains by multiplying the gas surface density by a factor of a thousandth.
        
        The difference between the two models is the assumed equilibrium temperature of the central planet, so the isothermal temperature profiles adopted differ. In each model, we incorporated four moons (each with a mass of 0.02 $M_\textrm{J}$) embedded in the CPMOD on resonant orbits, while a one-Solar-mass star is orbiting around the planet and the CPMOD at a 5 au distance. Because in hydrodynamic simulations, gravitational effects of moons are introduced gradually after the disk is fully relaxed. We thus can obtain two full disk models (a fully relaxed disk without moons) at no cost. 
        
        Then we run radiative transfer simulations to obtain the dust temperature, which shows good agreement with the temperature profiles used in hydrodynamic simulations (Figure \ref{fig:T_r}). We further calculate their SEDs following the same approach. The technical details of hydrodynamic models are described in Appendix \ref{appendix:hddetail}.

    \begin{figure*}[t!]
        \centering
        \includegraphics[width=0.95\textwidth]{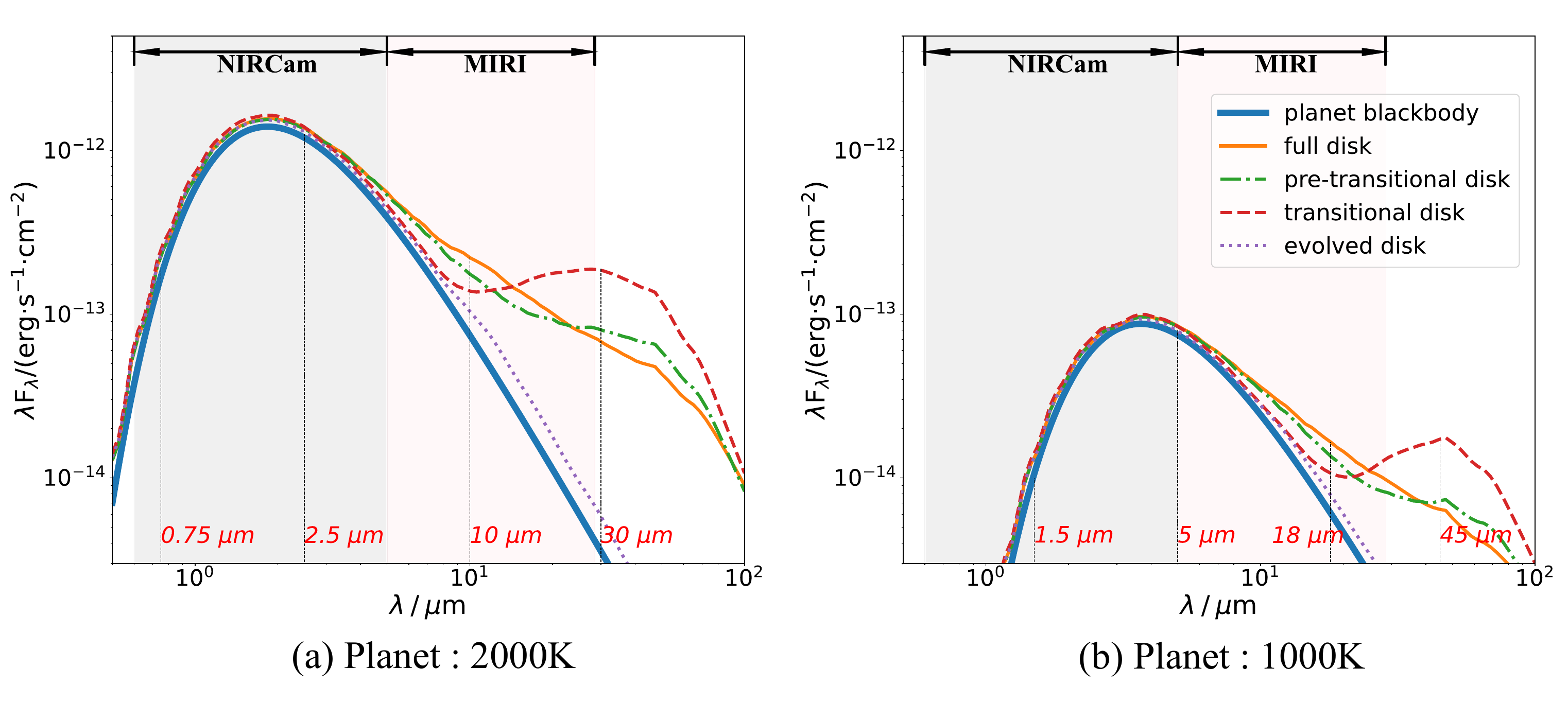}
        \caption{The SED calculated by radiative transfer simulations of four CPMOD prototypes with a central planet at 2,000 K (a) and 1,000 K (b), respectively. The horizontal axis represents the wavelength, ranging from 0.5 to 100 $\mathrm{\mu m}$, and the vertical axis is $\lambda  F_\lambda$, both in logarithmic scale. The thick blue solid lines indicate black body radiations from central planets. The thin orange solid, green dash-dotted, red dashed and purple dotted lines represent full, pre-transitional, transitional and evolved disks, respectively. The gray and pink shaded areas in the figure indicate the wavelength coverage of NIRCam and MIRI filters, respectively.}
        \label{fig:SED}
    \end{figure*}
    
\section{Results and Discussions} \label{sec:result}
    
    In this section, we present the results and analysis of our radiative transfer simulations for four prototypes of CPMODs. We further discuss the implication, namely classification of CPMODs based on their SEDs.
        
    \subsection{Temperature Profiles of Four Prototypes of CPMODs} \label{sec:temp} 
        
        We show the temperature distributions of four prototypes of CPMODs obtained from radiative transfer simulations in the right column of Figure \ref{fig:model}.

        For the full disk prototype, the temperature profile varies slowly with radius. For most part of the disk, temperature near the midplane is relatively cold, while above the mid-plane temperature increases significantly. The whole disk near the midplane is optically thick.

        For the pre-transitional disk prototype, the gap is optically thin, and a warm "wall" that is optically thick forms near the inner edge of the outer disk. As a result, the interior to the wall gets warmed up as shown in the temperature profile plot of the pre-transitional disk in Figure \ref{fig:model}. Towards large radius, the difference between the pre-transitional disk and the full disk becomes marginal.

        The transitional disk prototype exhibits elevated temperatures within its cavity as a result of optical thinness. The outer disk, exposed to more radiation, registers higher temperatures. Notably, the optically thin cavity results in a warm wall along the inner edge of the outer disk.

        Considering the low density throughout the evolved disk, the entire disk is optically thin, resulting in a higher temperature. However, because the total dust mass is low, the overall radiation is markedly weak for the evolved disk.
    
    \subsection{SEDs of Four CPMOD Prototypes} \label{sec:SED}
         
        In this subsection, we continue to explore the impact of structural differences between CPMOD prototypes on their SEDs. Our approach is similar to that applied in the analysis of PPD SEDs \citep{2014prpl.conf..497E}. For illustration purposes, in Figure \ref{fig:SED} we show SEDs from four CPMOD prototypes around a 10 $M_\mathrm{J}$ planet with an equilibrium temperature at 1,000 and 2,000 K, respectively. In this work, we have assumed that the radiation from the planet is a pure blackbody. Because the central object is much cooler than a star, the peak of the overall radiation peaks at near-infrared wavelengths.
        
        For the case of a planet at 2,000 K, as shown in panel a of Fig. \ref{fig:SED}, the radiation in the near infrared band ($0.75 \sim 2.5\: \mathrm{\mu m}$) is mainly contributed by the central object. There is little difference between the four types of disks in this band.

        On the other hand, radiation in the $2.5 \sim 10 \: \mathrm{\mu m}$ band is primarily provided by the inner disk, with the full disk contributing the most radiation, resulting in the SED curve at the top. The pre-transitional disk follows with slightly lower radiation than the full disk. The internal regions of both transitional and evolved disks, being optically thin, contribute the least in this band, reflected by the SED curve at the bottom.
        
        The abrupt increase in the SED curve of the transitional disk in the $10 \sim 30 \: \mathrm{\mu m}$ band is attributed to the presence of a warm wall -- the inner edge of a warmer outer disk (see the temperature profile of the transitional disk in Figure \ref{fig:model} and the discussion in Section \ref{sec:temp}), which dominates the radiation in this band.
        
        Above 30 $\mathrm{\mu m}$, radiation is mainly contributed by relatively cold outer disks. Evolved disks are generally tenuous and optically thin, so the emission of outer disks is very low, and the radiation curve decreases rapidly. 
        
        For the central planets with a lower temperature at 1,000 K, the trends discussed above still hold, but the peak blackbody radiation and the overall contribution of low-temperature disks shift toward the longer wavelengths, as illustrated in panel b of Fig. \ref{fig:SED}.

        Here we show that an SED of a CPMOD can be broken down into a blackbody radiation from the planet and a dust disk emission. In principle, fundamental information about a real CPMOD can be revealed by comparing its spectrum with SEDs derived from four prototypes of CPMODs. However, in practice, there are a large number of CPMOD candidates, and obtaining a full spectrum for each target would be resource intensive. A more efficient and convenient approach involving fewer observational efforts is desirable. 
    
    \subsection{Implications for Color-color Diagrams}\label{sec:color}
        In light of distinct characteristics of SEDs presented above, we utilize the color-color diagram method \citep{1992ApJ...393..278L, Hernandez_2007}, which has been successfully applied to \textit{Spitzer} data to classify various types of protoplanetary disks \citep[e.g., see the review by][]{2014prpl.conf..497E}. 
    
        \begin{figure*}[t!]
            \centering
            \includegraphics[width=0.65\linewidth]{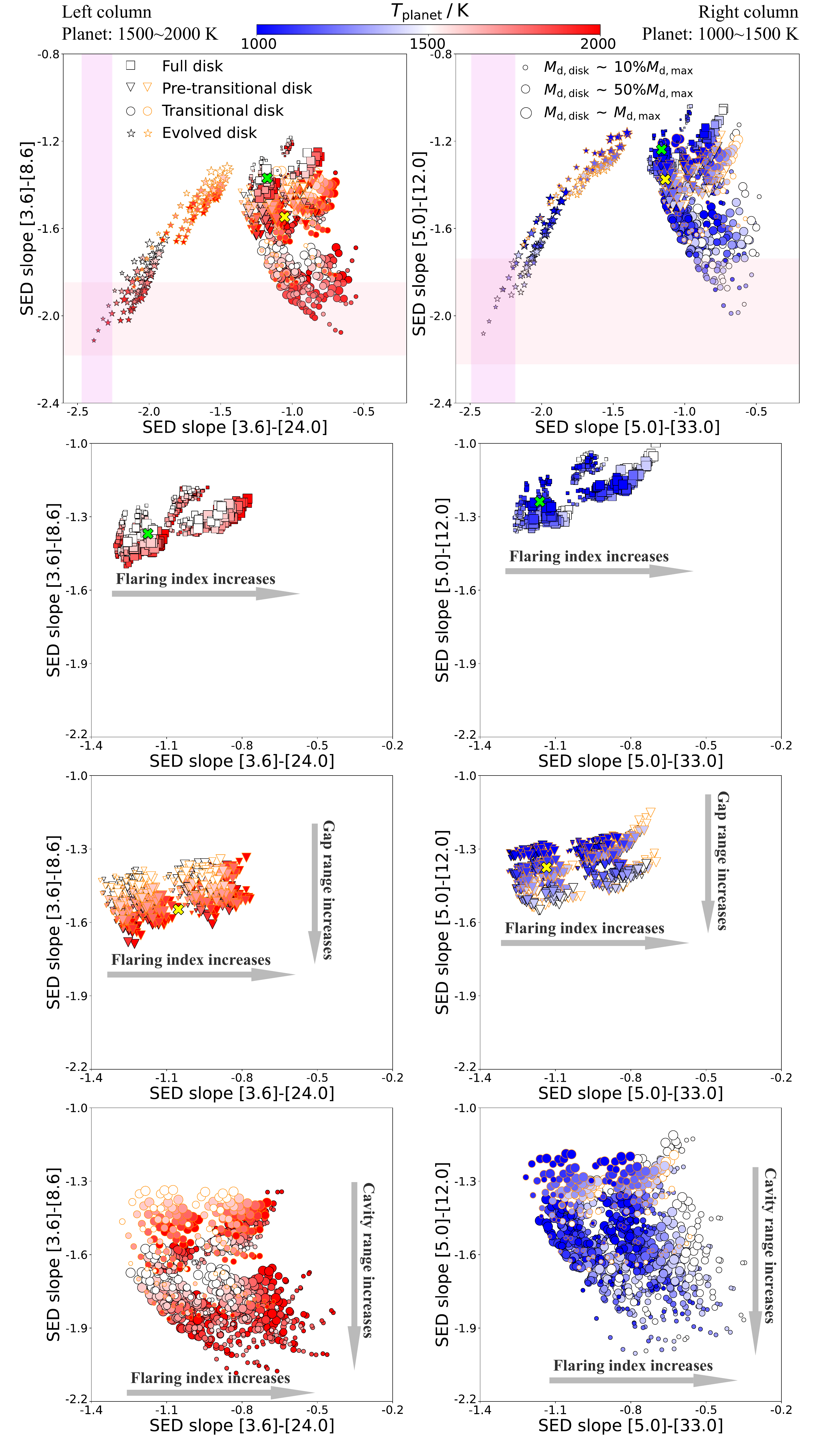}
            \caption{
            SED slope color-color diagrams of various parametric models. Selected four types of disk models are divided into two subplots on the first row by the temperature of central object (1,500$\sim$2,000 K and 1,000$\sim$1,500 K) for illustration. Different shaped points represent different types of models. \textbf{Square}: full disks, \textbf{Inverted triangle}: pre-transitional disks, \textbf{Circle}: transitional disks, \textbf{Pentagram}: evolved disks. Symbols are color-coded to indicate the temperature of the central planet, ranging from 1,000 to 2,000 K. The size of the symbol represents different disk density values, with larger points indicating that they have higher density in their own type (i.e., the larger dust mass of the disk). For the later three types of disks, black and orange edge colors indicate depletion factors of $10^{-4}$ and $10^{-3}$, respectively. The shadowed regions in the two top panels correspond to colors of planet blackbody radiation with a temperature within 1,500$\sim$2,000 K and 1,000$\sim$1,500 K, respectively. In addition, the crosses appear in diagrams on the left and right columns represent results of hydrodynamic models with a host temperature at 2,000 K and 1,000 K respectively, with green representing the full disk and yellow representing the pre-transitional disk. We also show all models of full, pre-transitional and transitional disks on each row for comparison.}
            \label{fig:color-color-diff}
        \end{figure*}

        We define the SED slope (or color) as
        \begin{equation}\label{eq:12}
            [\lambda_1]-[\lambda_2] = \frac{\Delta \log(\lambda F_{\lambda} )}{\Delta \log(\lambda)} = \frac{\log(\lambda_1 F_{\lambda_1})-\log(\lambda_2 F_{\lambda_2})}{\log(\lambda_1)-\log(\lambda_2)},
        \end{equation}
        where $\lambda$ represents wavelength points, and $F_{\lambda}$ is the flux per unit wavelength. Subscripts 1 and 2 indicate two different bands. Note that in some studies \citep{Furlan_2006}, it is called the ``spectral index $n$''.

        Here $\lambda_1$ is chosen as a reference point such as wavelengths of 3.6 or 4.5 $\mathrm{\mu m}$ (\textit{Spitzer IRAC} bands) at which the blackbody radiation from the planet always dominates the disk emission. Obviously, slopes determined by other wavelengths ($\lambda_2$ and $\lambda_3$) should reflect the difference of SEDs between different types of CPMOD.

        For illustrative purposes, we first plot all our CPMOD models on color-color diagrams (Figure \ref{fig:color-color-diff}) in which we separate the samples into two subgroups based on the temperature of the host planet. For CPMODs around hot planets, we adopted a reference wavelength $\lambda_1$ of 3.6 $\mu$m, and the other two wavelengths are $\lambda_2 = 8.6\; \mu$m and $\lambda_3 = 24.0\; \mu$m, respectively. Each of the shadowed regions corresponds to a color ($[\lambda_1]-[\lambda_2]$ or $[\lambda_1]-[\lambda_3]$) of the black-body radiation of the planet. A black body SED is thus confined in the lower-left rectangular region on the color-color diagram. The more emission a disk irradiates at these wavelengths, the further it deviates from the shadowed regions on the color-color diagram. Each type of CPMODs has distinct colors, and therefore CPMODs are clustered in certain regions on the plot making them potentially distinguishable.

        Similar to previous classification results for PPDs, full disks reside in the upper right region with respect to other types (see the top row of Figure \ref{fig:color-color-diff}), simply because a full disk differs from a pure black body in both colors (compare SED slopes we defined in Equation \ref{eq:12} between the blue and orange solid lines in Figure \ref{fig:SED}). Obviously, the position of a full disk on such a diagram depends on: (1) host temperature, which determines the black-body SED of the host; and (2) dust mass, which affects how flux is intercepted and reprocessed. Besides, disk geometry plays an important role. We show there is a spread due to changes of disk flaring index in the second row of Figure \ref{fig:color-color-diff}. This is because in a more flaring disk it is easier to receive host irradiation at large radius.
       
        Evolved disks are mainly concentrated in the lower left region. Not surprisingly, the least massive evolved disks are too tenuous to intercept and reprocess host flux. Consequently, their colors are almost identical to a pure blackbody. When the overall depletion factor increases from $10^{-4}$ to $10^{-3}$, an evolved disk moves towards upper right, as the disk becomes more opaque and its color becomes closer to that of a full disk.
       
        Transitional disks cover a large parameter space in the lower-right region on the color-color diagram (the bottom row of Figure \ref{fig:color-color-diff}). Except for the obvious causes for such a spread, i.e. host temperature and dust mass, disk cavity has a major impact. On one hand, a tenuous disk cavity is similar to the inner part of a tenuous evolved disk, so they share a similar color defined by the vertical axis of color-color diagram. However, a shallower cavity, e.g. an increase of depletion factor from $10^{-4}$ to $10^{-3}$, can cause considerable degeneracy in color-color classification. On the other hand, transitional disks with wide cavities are easier to be distinguished from other types on color-color diagram. As for the horizontal spread, geometry of outer disk is responsible for the the same reason that we have explained in the full disk case.
       
        Pre-transitional disks are difficult to be separated from full disks on color-color diagram. In general, the points representing pre-transitional disks will be positioned below those representing full disks. In reality, however, due to all possible gap depletion level, gap width and geometry of outer disk as well as host irradiation and dust mass, even pre-transitional disks cannot be distinguished from transitional disks as long as the depletion of the inner disk becomes moderate.

        For CPMODs around cool planets, we adopt slightly longer wavelengths ($\lambda_1 = 5.0\; \mu$m, $\lambda_2 = 12.0\; \mu$m and $\lambda_3 = 33.0\; \mu$m) for better performance, as their SEDs are all red-shifted (compare the two panels of Figure \ref{fig:SED}). The clustering trend is similar to the subgroup around hot planets.

        Here, we summarize a set of empirical formulas to guide the selection of suitable wavelength points:
            \begin{equation}\label{eq:13}
                \lambda _1 = \frac{b^{\prime}}{T} \times \gamma_1,
            \end{equation}
            \begin{equation}\label{eq:14}
                \lambda _2 = \frac{b^{\prime}}{T} \times \gamma_2,
            \end{equation}
            \begin{equation}\label{eq:15}
                \lambda _3 = \frac{b^{\prime}}{T} \times \gamma_3,
            \end{equation}
        where $\lambda_1$, $\lambda_2$, and $\lambda_3$ are the wavelength points in [$\lambda_1$]-[$\lambda_2$] and [$\lambda_1$]-[$\lambda_3$]. $b^{\prime}$ represents a constant similar to Wien’s constant $b$ in Wien's displacement law $\lambda_\mathrm{max}=b/T$, and it follows $\lambda_\mathrm{max}^{\prime}=b^{\prime}/T$, where $\lambda_\mathrm{max}^{\prime}$ is the wavelength at the peak of $\lambda F_{\lambda}$ rather than $F_{\lambda}$. Similar to Wien's law, based on the derivation and numerical calculation, $b^{\prime}\approx3674\: \mathrm{\mu m\cdot K}$ (see the derivation in Appendix \ref{appendix:deriv}). 
        $T$ stands for the adopted temperature for the central planet. 
        $\gamma_1=1.7$, $\gamma_2=4.1$, and $\gamma_3=11.4$ are three empirical coefficients. With these empirical formulas, one can adjust three wavelength points for different targets accordingly.
 
        \begin{figure*}[t!]
            \centering
            \includegraphics[width=0.95\textwidth]{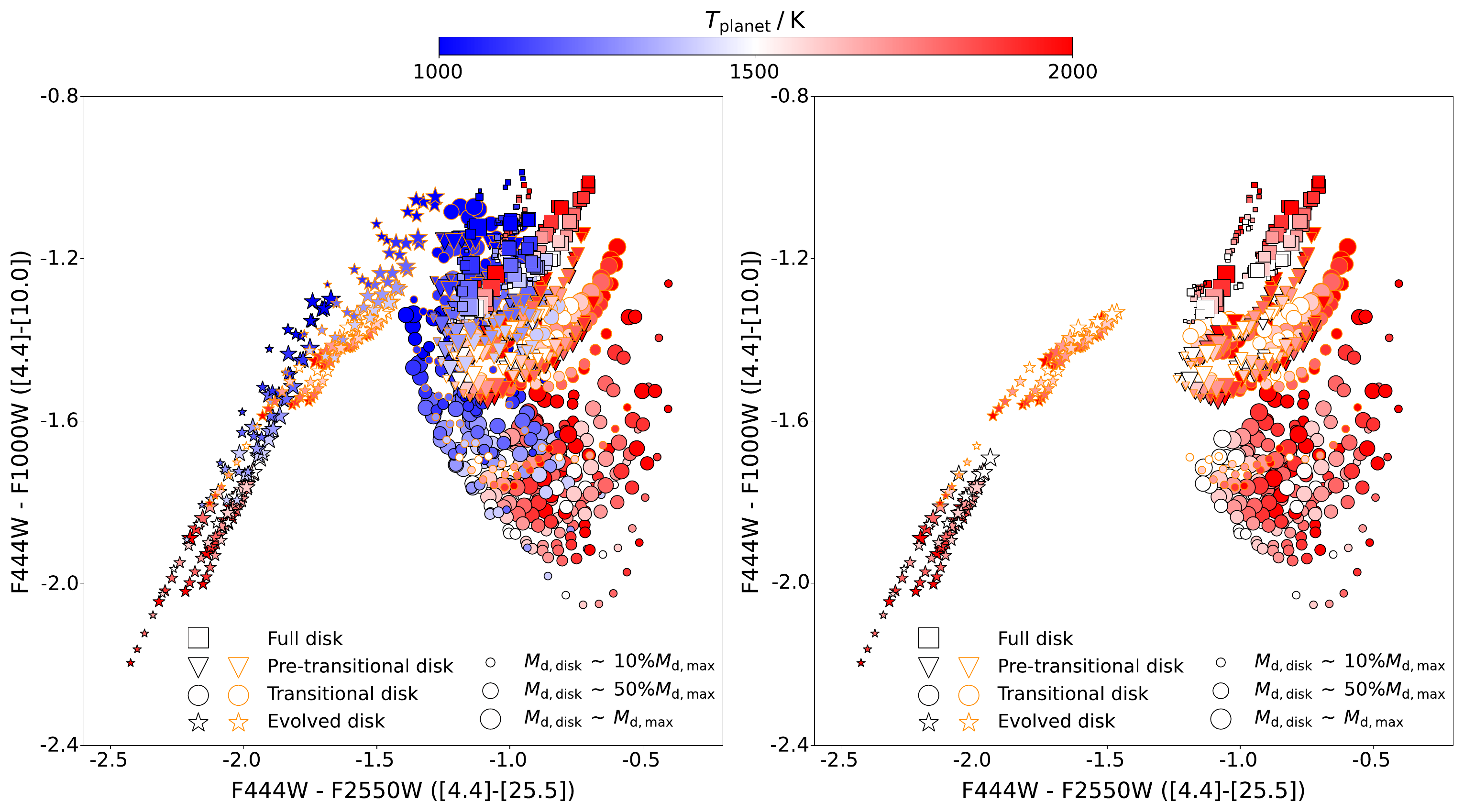}
            \caption{Color-color diagram of CMPOD models around hot and cool hosts (left) and around hot host only (right) in JWST filters. Symbols are same as Figure 3. About a quarter of our models are randomly picked to make the scatter plot.
            }
            \label{fig:color-color}
        \end{figure*}
       
        By measuring radiation flux at only three wavelengths, we demonstrate that the color-color diagram is clearly a more efficient classification method than comparing different SEDs directly. As for the classification performance, color-color diagram is almost the same with comparing SEDs because when two disks are degenerate in their colors, their SEDs also look similar. However, there is a caveat that one needs to bear in mind, i.e. in real observation we may not have a priori knowledge of the planet's temperature. The color-color difference between the SED models could be more degenerate if the selection of three wavelengths is not ideal. Besides, a realistic color is obtained through comparing photometry with infrared bandpass filters. We will discuss the impact of NIRCam and MIRI wide filters and how we adapt the color-color classification scheme.        
            
    \subsection{Observational Strategies with Synthesized JWST Color-color Diagram}

        Considering JWST/MIRI imager's coverage in mid-infrared from 5.6 to 25.5 $\mu$m 
        plus a near-infrared wavelength coverage from 0.6 to 5.0 $\mu $m by JWST/NIRCam, the utilization of color-color diagram is more suitable to CPMODs around hot planets according to previous analysis, which can be understood in the sense that MIRI fails to cover the peak radiation from a cold CPMOD in mid-infrared (see the right panel of Figure \ref{fig:SED}).
        Therefore, given a sample of candidate CPMODs to classify, it is ideal to exclude sources around cool hosts (say planets with a temperature lower than 1,500 K) if host temperature can be constrained.

        Adapting our classification for JWST observations, we calculate synthetic colors of all our models with NIRCam and MIRI filters. Here we use three wide filters, F444W of NIRCam, F1000W and F2550W of MIRI, with pivot wavelengths at 4.4, 10.0 and 25.5 $\mu$m to define synthesized colors, respectively. 
        
        The calculation of an SED slope by Equation \ref{eq:12} needs to be updated accordingly as the single wavelength point $\lambda$ is replaced by a pivot wavelength $\Bar{\lambda}$ with a bandwidth $\Delta \lambda$. 
        \begin{equation}\label{eq:filteredslope}
            [\overline{\lambda_1}]-[\overline{\lambda_2}] = \frac{\Delta \log(\overline{\lambda F_{\lambda}} )}{\Delta \log(\Bar{\lambda})}  = \frac{\log(\overline{\lambda_1 F_{\lambda_1}})-\log(\overline{\lambda_2 F_{\lambda_2}})}{\log(\overline{\lambda_1})-\log(\overline{\lambda_2})},
        \end{equation}
        where $\overline{\lambda F_{\lambda}}$ represents the flux per unit wavelength times wavelength, filtered with the filters' transmission curves.
        
        We find that the filter bandpass does not affect our color-color classification because bandwidths $\Delta \lambda$ of those wide filter are only 1.024, 1.80 and 3.67 $\mu$m, respectively.
 
        In the left panel of Figure \ref{fig:color-color}, we plot random selected models including those with cool hosts on the synthetic color-color diagram. Because the planet temperature could vary within a wide range between 1,000 and 2,000 K, the classification efficiency becomes inferior. However, if one can preclude the degeneracy caused by cool sources, CPMODs with host temperature above 1,500 K can be effectively classified with synthesized JWST colors (see the right panel of Figure \ref{fig:color-color}).
        
        The clustering tread of each type of CPMOD has been thoroughly discussed in Section \ref{sec:color}. Within the parameter space we have explored (Table \ref{tab:model}), we find that our synthetic color-color classification can distinguish different CPMOD models around hot planets ($ 1,500\;\mathrm{K} \leq T_\mathrm{eff} \leq 2,000\;\mathrm{K}$). In particular, evolved disks are easiest to separate from other types of disk. A transitional disk is distinguishable if its cavity is wide and/or severely depleted in dust. Pre-transitional disks are sometimes difficult to identify especially when their inner gaps are not optically thin enough.

        We also note that our synthesized JWST color-color diagram is not ideal to classify CPMODs around cool planets ($ 1,000\;\mathrm{K} \leq T_\mathrm{eff} \leq 1,500\;\mathrm{K}$) simply because the peak emission of their disk components cannot be captured by MIRI. If one could extend the wavelength coverage up to 45 $\mu m$, CPMODs around cool targets could also be classified with a similar manner.   

        In summary, our goal is to provide reference models to show what could be done with near-to-mid-infrared photometry for ongoing and future JWST studies of CPMODs. Given the vast parameter space not covered in this study, tailored modeling efforts for individual systems would be needed for more realistic and specific comparisons.

    \subsection{Impacts from Other Parameters}            

        We discuss some other factors that could influence the color-color classification in this subsection. For instance, we only consider a fixed planet mass throughout our models for illustration. This may have an impact on the dust depletion in gaps and cavities. We have tested two levels of dust depletion in our models.
        
        The black body SED is determined by the equilibrium temperature and physical size of a planet. The former one affects SED's peak, while the latter one affects the total amount of photons. In other word, changing planet temperature will cause the SED to move horizontally, and changing planet size will cause the SED to move vertically. The temperature variation has been studied in our model. Small planetary size variation does not change the shape of an SED, thus the synthesized color remains the same.

        We do not consider the influence of viewing angle. As long as the inclination angle $i$ is less than $70^\circ$ ($i = 0^\circ$ is face-on and $i = 90^\circ$ is edge-on), the impact on SED is negligible \citep{Whitney_2003}. So our color-color classification can be applied to most of the disks.

        Polycyclic aromatic hydrocarbons (PAHs) are ubiquitous complex molecules found in the interstellar medium and are commonly associated with star formation \citep{2004ApJ...613..986P}. Their broad emissions at 3.3, 6.2, 7.7, 8.6, 11.3 and 12.7 $\mu m$ can potentially contaminate the classification if exist. In this study, The filters we chosen have avoided those wavelength ranges. And there are specific filters ideal for photometry in those bands. Besides, in a recent observation of GQ Lup B \citep{cugno2024midinfrared}, no emission feature near 11.3 $\mu m$ was found.

        Last but not the least, we assume a steady state grain size distribution in our modeling. As dusts undergo coagulation and fragmentation, grain size distribution evolves as well \citep[see, e.g.][]{Meru_2013} Grain growths may eventually lead to moon formation, which could open gaps in disks like pre-transitional disks. Such structures in disk could trap dusts and make the situation even more complicated \citep{2018ApJ...857...87L}. Obviously, such a topic is out of the scope of this paper and deserves more future studies.

    \subsection{Results of Hydrodynamic Simulations} \label{sec:hdresult}

        In addition, we validate our parametric models of full and pre-transitional disks with two hydrodynamic simulations (see the detail laid out in Appendix \ref{sec:hd-simulation}), based on which we further conducted radiative transfer simulations. These results are plotted as cross markers in the color-color diagram (Figure \ref{fig:color-color-diff}).
        
        Despite the gap depths in the hydrodynamic model being slightly shallower than those adopted in parametric models ($10^{-4}$ and $10^{-3}$), their colors are consistent with those of parametric full and pre-transitional disks. For the pre-transitional disk in our hydrodynamic simulation, a depletion factor of $10^{-2}$ in a wide gap is already distinct from a full disk in near-to-mid-infrared colors. We also notice that gaps formed through satellite-disk interaction are not uniformly depleted throughout. Gap width measured from outside the gap is much wider than that measured from the bottom, and density bumps near gap edges also affect the direct comparison with parametric models \citep{2018ApJ...857...87L}.
        
        Nonetheless, such an agreement not only verifies our parametric models, but also inspires us the possibility of studying the substructure in CPMODs. For instance, the formation of moons may open a gap in a full CPMOD, so a full disk can evolve into a pre-transitional disk. With the help of future CPMOD observations, we may gain more insights into the subject of moon formation and CPMOD evolution.
    
\section{Conclusions} \label{sec:conclusion}

    We analytically constructed a series 3D disk models representing CPMODs around a ten Jovian-mass planet at different evolutionary stages, i.e. full, pre-transitional, transitional, evolved circum-planetary-mass-object disks, which are similar to the circumstellar disk counterparts. The range of total dust mass of full disk models is between $2 \times 10^{-7}$ and $1 \times 10^{-4}$ Jovian masses. And about one tenth of the total dust mass, i.e. $2 \times 10^{-8}$ and $1 \times 10^{-5}$, is locked in small dust grains ranging from 0.1 to 10 $\mu m$. In addition, Our analytic full and pre-transitional CPMOD models can be reasonably reproduced by 2D hydrodynamic simulations, in which gap formation is through satellite-disk interactions. 
    
    Assuming an equilibrium temperature between 1,000 and 2,000 K for the central planet and a dust size distribution of 0.1 to 10 $\mu$m, we calculate the temperature profile for each disk model using Monte Carlo simulations accordingly. Then we conduct radiative transfer simulations to obtain its SED. In the end, we build an SED ensemble of 4,624 CPMOD models covering a wide parameter space including four models constructed based on hydrodynamic simulations. The key conclusions of this study are:

    \begin{itemize}
        \item 
        We propose an empirical color-color classification scheme and apply it to various circum-planetary-mass-object disk models built upon key parameters. We show four presumable types of CPMODs, i.e. full, pre-transitional, transitional and evolved circum-planetary-mass-object disks reside in different parameter space on color-color diagram.
        \item
        Specifically, we use photometry for three wide filters, F444W of NIRCam, F1000W and F2550W of MIRI, to compute F444W - F1000W and F444W - F2550W colors for our CPMOD models. We demonstrate that synthesized color-color diagram is ideal to distinguish various disk models around hot planets ($ 1,500\;\mathrm{K} \leq T_\mathrm{eff} \leq 2,000\;\mathrm{K}$). 
        \item 
        For CPMODs around cool planets ($ 1,000\;\mathrm{K} \leq T_\mathrm{eff} \leq 1,500\;\mathrm{K}$), an effective color-color classification would require filters at longer wavelength (e.g. $\sim 45\; \mu m$), which are not available at the moment. 
 
    \end{itemize}

    Among the targets of JWST cycle 3 programs, spectra ($1 \sim 14 \: \mathrm{\mu m}$) of 8 free floating planetary-mass objects (FFPMOs) with disk structures will be observed (JWST proposal 4583), and a NIR-MIR spectrum ($0.97 \sim 27.9 \: \mathrm{\mu m}$) of a planetary-mass companion SR 12 c and its disk will be characterized (JWST proposal 6086). We hope our CPMOD classification scheme could benefit from these ongoing programs and inform us the true nature of satellite-forming structures around young planetary objects.

\section*{Acknowledgement}
    We thank the referee for many insightful comments that greatly improve the quality of this paper. We express our gratitude to Yinhao Wu for his assistance and valuable suggestions throughout this project. We appreciate the advice and support provided by Xiaoyi Ma in radiative transfer simulations. The authors also thank Zhizhen Qin for his assistance in the development of hydrodynamic codes and numerical simulations. We also thank Yixian Chen, Yanfei Jiang, Yaping Li, Doug Lin, Xing Wei and Yong Zhang for their insightful discussions. Part of the simulations in this study utilized the resources of ``graham'', a heterogeneous cluster supercomputer provided by The Digital Research Alliance of Canada.  S.-F. Liu acknowledges the support from the Guangdong Basic and Applied Basic Research Foundation under grant No. 2021B1515020090,  the National Natural Science Foundation of China under grant No. 11903089, and the China Manned Space Project under grant Nos. CMS-CSST-2021-A11 and CMS-CSST-2021-B09.  P. Huang acknowledges the financial support from the Canadian funding NSERC ALLRP 577027-22. X. Sun would like to thank the hospitality of the Protoplanetary Disk and Planet Formation Summer School organized by Xuening Bai and Ruobing Dong in 2022, hosted by the Chinese Center for Advanced Science and Technology.

%

\vspace{5mm}
\facilities{JWST (NIRCam and MIRI)}

\software{\texttt{RADMC-3D} \citep{2012ascl.soft02015D},  
          \texttt{Athena++} \citep{2020ApJS..249....4S},
          \texttt{REBOUND} \citep{2012A&A...537A.128R},
          \texttt{dsharp\_opac} \citep{Birnstiel_2018}
          }



\appendix

\section{Dust Opacity}\label{appendix:Dustopac}

    \begin{figure*}[ht!]
        \centering
        \includegraphics[width=0.95\textwidth]{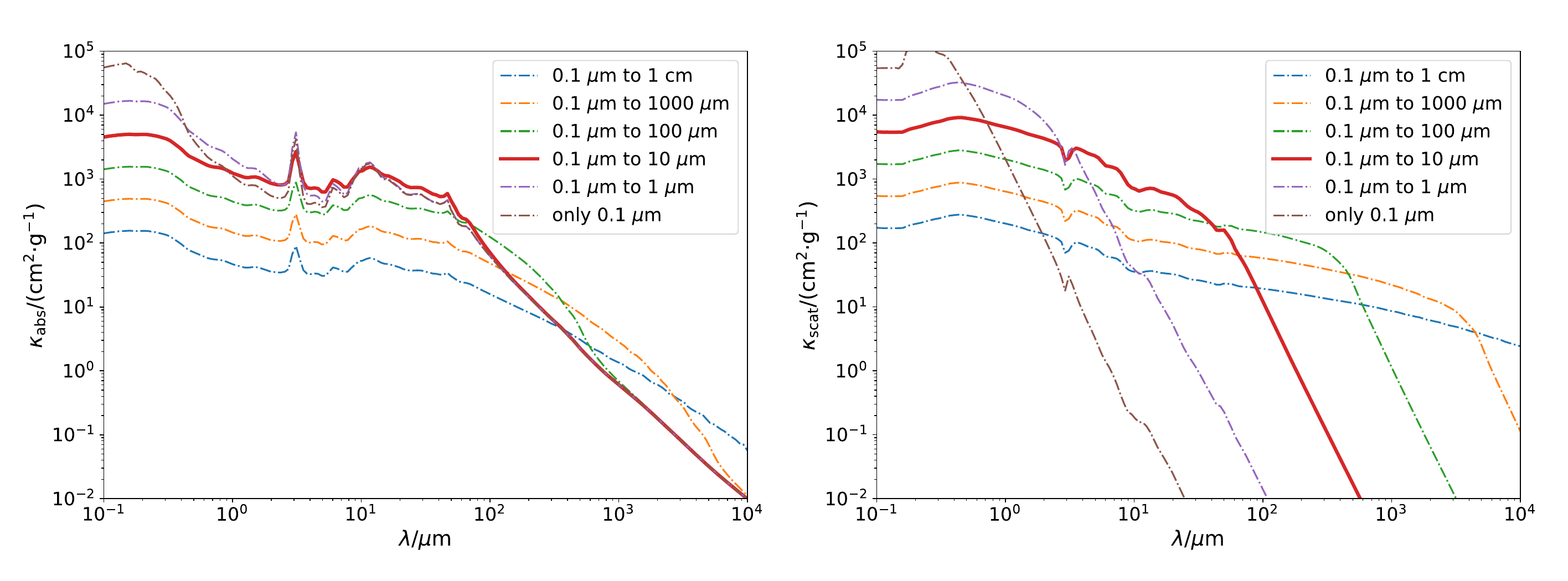}
        \caption{
            The averaged dust opacity curves for absorption(left) and scatting(right) of different size distributions, show the opacity as a function of wavelength. The solid line represents the dust grain size ranging from 0.1 to 10 micron (the one we adopted in our modeling), and the dash-dot lines show the opacity curves of other size distributions for reference.
            }
        \label{fig:opac}
    \end{figure*}

    We use \texttt{dsharp\_opac} to obtain the averaged dust opacity curves for absorption and scatting of different size distributions, as shown in Figure \ref{fig:opac}. All of these curves follow a power-law size distribution $n(a) \propto a^{-q}$, where the power index $q=3.5$. The solid line represents the dust grain size ranging from 0.1 to 10 micron, which is used in this work. The dash-dot lines show the opacity curves of other size distributions are also plotted for reference.
    
\section{Hydrodynamic Simulation}\label{appendix:hddetail}

    We first introduce the equations in Section \ref{sec:hd-equations}, then continue with a summary of the key features of the numerical method in Section \ref{sec:hd-conditions} \& \ref{sec:hd-simulation}.
        
        \subsection{Equations}\label{sec:hd-equations}
    
            The continuity and momentum equations for this CPMOD simulation are as follows~\citep{Zhang2024}:
            \begin{equation}
                \frac{\partial \Sigma}{\partial t}+\nabla \cdot(\Sigma \boldsymbol{v})=0,
            \end{equation}
            \begin{equation}
                \frac{\partial(\Sigma \boldsymbol{v})}{\partial t}+\nabla \cdot(\Sigma \boldsymbol{v} \boldsymbol{v}+P \mathsf{I})=-\Sigma \nabla \Phi-\nabla \cdot \mathsf{T}_{\mathrm{vis}},
            \end{equation}
            
            In the above equations, $\boldsymbol{v}$ is the gas velocities, $P$ is the vertically integrated pressure, $\mathsf{I}$ is the identity tensor and $\mathsf{T}$ is the viscous tensor:
            \begin{equation}
                \mathsf{T}_{\mathrm{vis}, i j}=-\Sigma \nu\left(\frac{\partial v_i}{\partial x_j}+\frac{\partial v_j}{\partial x_i}-\frac{2}{3} \delta_{i j} \nabla \cdot \boldsymbol{v}\right) .
            \end{equation}
    
            We use the locally isothermal equation of state in this study:
            \begin{equation}
                P = c_\mathrm{s}^2 \Sigma,
            \end{equation}
            where $P$ is the vertically integrated pressure
    
            In a polar coordinate system (radius r, azimuth $\phi$) centered on the planet, the potential $\Phi$ is
            \begin{equation}\label{eq:5}
                \Phi=-\frac{GM_\mathrm{p}}{r}+\Phi_{\ast}+\sum_\mathrm{i=1}^{N_\mathrm{m}} \Phi_\mathrm{m,i},
            \end{equation}
            \begin{equation}\label{eq:6}
                \Phi_{\ast}=-\frac{GM_{\ast}}{r_{\ast}}+\frac{GM_{\ast}r \cos \phi_{\ast}^{\prime}}{r_{\ast}^2},
            \end{equation}
            \begin{equation}\label{eq:7}
                \begin{split}
                \Phi_\mathrm{m,i}=&-\frac{GM_\mathrm{m,i}}{\sqrt{r^2+r^2_\mathrm{m,i}-2rr_\mathrm{m,i}\cos\phi_\mathrm{m,i}^{\prime}+r^2_\mathrm{s,i}}}
                \\&+\frac{GM_\mathrm{m,i}r \cos \phi_\mathrm{m,i}^{\prime}}{r_\mathrm{m,i}^2},   
                \end{split}
            \end{equation}
            where $G$ is the gravitational constant, $M_\mathrm{p}=10M_\mathrm{J}$ is the planet mass, $N_m=4$ is the total number of moons, $\Phi_{\ast}$ and $\Phi_\mathrm{m}$is the star and moon's gravitational potential, $M_{\ast}$ and $M_\mathrm{m}$ represents the mass of star and moons, $r_{\ast}$ and $r_m$ the star and moon's radial coordinate, $r_s$ the smoothing length of the moon's potential, $\phi_{\ast}^{\prime}=\phi-\phi_{\ast}$ the azimuthal separation from the star and $\phi_\mathrm{m}^{\prime}=\phi-\phi_\mathrm{m}$ the azimuthal separation from the moon. The latter term of Equation \ref{eq:6} and Equation \ref{eq:7} is the indirect term due to our frame centering on the planet rather than the center of mass. 
        
        \subsection{Initial Conditions} \label{sec:hd-conditions}
        
            The inner and outer boundaries are set at 5 and 500 $R_\mathrm{J}$ (approximately 0.003 to 0.3 $R_\mathrm{H}$), respectively, where $R_\mathrm{H}$ represents the Hill radius of our central object-a planet with a mass equivalent to 10 Jupiter masses. The choice of 0.3 $R_\mathrm{H}$ for the outer boundary is motivated by the truncation of gas in the CPDs, occurring at approximately one-third of the Hill radius \citep{Benisty_2021}. On the outer side of the disk, akin to the Sun-Jupiter system, we positioned a solar-mass star at a distance of 5.2 au from the planet in the CPD. So the Hill radius is
            \begin{equation}\label{eq:4}
                R_\mathrm{H}=5.2\cdot\sqrt[3]{\frac{1}{3} \cdot \frac{10M_\mathrm{J}}{M_\mathrm{{\odot}}}}\:\mathrm{au}\approx1636.7\:R_\mathrm{J}.
            \end{equation}
    
            We put a total of 4 moons into the model, orbiting in a 1:2:4:8 resonant pattern, and their semi-major axes are 12.7, 20.1, 32.2, and 51.0 $R_\mathrm{J}$, respectively. Initially, both the star and moons follow Keplerian orbits. Subsequently, we utilized a Runge-Kutta $4^{th}$ n-body code to compute their orbits. Given that the integral precision of the fluid part is second order, the $4^{th}$ n-body precision proves to be adequate.

            For gas density distribution, we refer to the previous work \citep{Fung_2019}, set $\gamma=-1.5$ and take the form of 
            \begin{equation}\label{eq:8}
                \Sigma_{g}(r)=\Sigma_{g,0} \cdot (r/r_0)^{-1.5}, 
            \end{equation}
            where $r_0=10\:R_\mathrm{J}$, same as it in the radiative transfer simulation, and $\Sigma_{g,0} = 10^3 \:\mathrm{g\cdot cm^{-2}}$.
            
            The kinematic viscosity in our simulation follows the formula of
            \begin{equation}\label{eq:9}
                \nu=\alpha c_\mathrm{s} h,
            \end{equation}
            where $\alpha$ is the Shakura-Sunyaev parameter \citep{1973A&A....24..337S}, $c_\mathrm{s}$ is the sound speed and $h$ the scale height. According to previous studies on the dead zone in CPDs \citep{Chen_2020}, the $\alpha$-parameter can reach $10^{-5}$ locally, so we choose a constant $\alpha=10^{-4}$ in our CPMOD model. 

            \begin{figure*}[t!]
                \centering
                \includegraphics[width=0.6\textwidth]{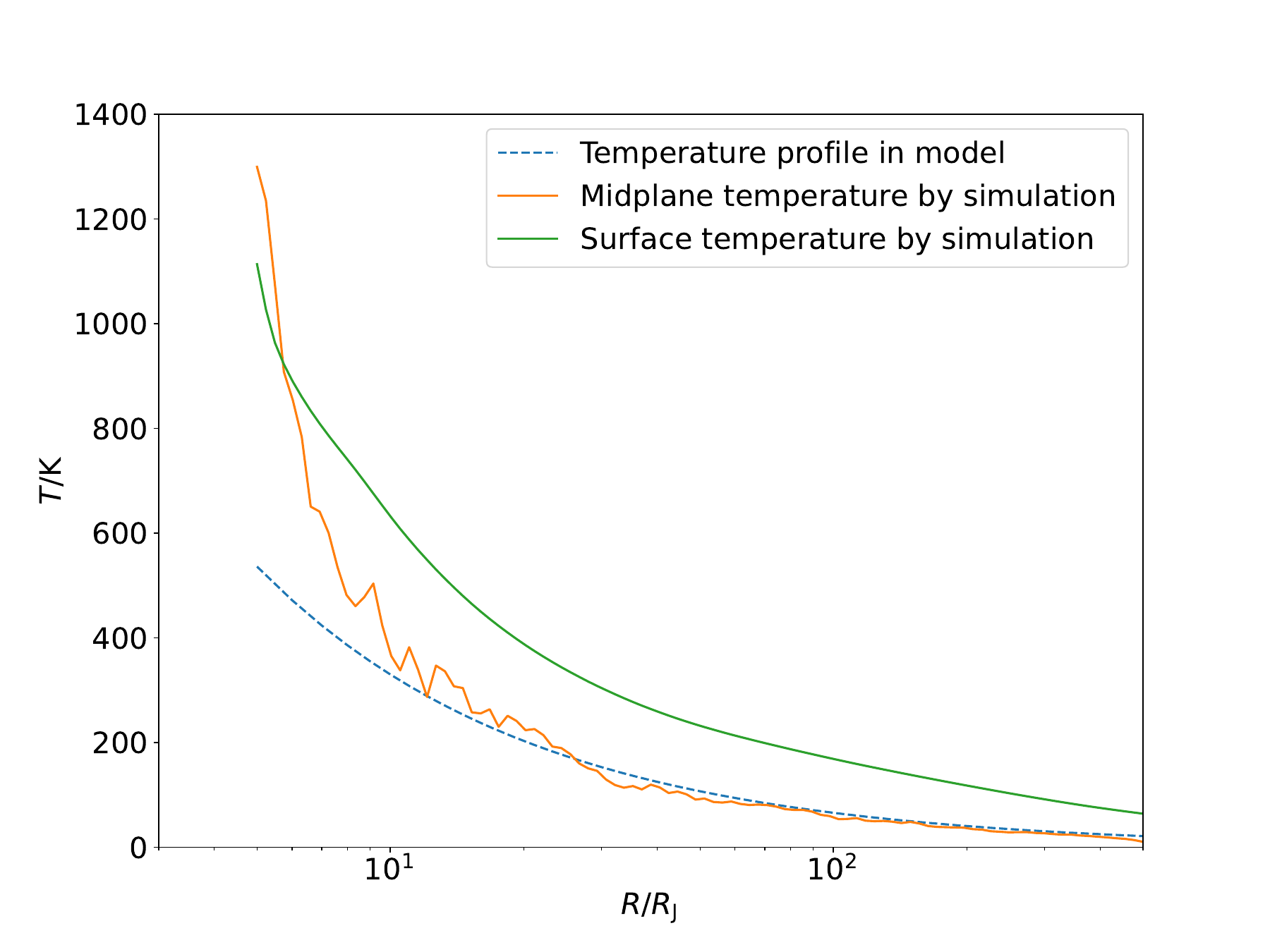}
                \caption{
                    The temperature profile of the 2,000K model, showing the distribution of temperature with radius. The solid lines are the results of radiative transfer simulation, while the orange one and the green one represent the mid-plane and the surface temperature, respectively. The dash line represents the mid-plane temperature profile used in the model. This is a comparison of the results after two iterations.
                }
                \label{fig:T_r}
            \end{figure*}
            
            We use an estimated temperature profile to run the preliminary simulation and plug the obtained hydrodynamic results into the radiative transfer simulation. Then, the mid-plane temperature distribution obtained by radiation transfer simulation is substituted back into the hydrodynamic simulation to ensure the self-consistency of the model. This process is iterated two times, as shown in Figure \ref{fig:T_r}. Our final mid-plane temperature distribution is 
            \begin{equation}\label{eq:2000K_tem}
                T(r)=330\cdot (r/r_0)^{-0.7}\:\mathrm{K}.
            \end{equation}
            Thus, the gas aspect ratio is
            \begin{equation}\label{eq:2000K_h/r}
                h/r= c_\mathrm{s}/v_{\mathrm{K}}=0.08\cdot (r/r_0)^{0.15},
            \end{equation}
            where $h$ is the scale height, $r$ is the radius, $v_{\mathrm{K}}$ represents the locally Keplerian velocity. Note that we have built two models in this section, and the Eq. \ref{eq:2000K_tem} and \ref{eq:2000K_h/r} here is only used to describe the first model, the central planet of which is 2,000 K; formulas of the other model needs to be corrected for the 1,000 K central planet, and follow:
            \begin{equation}\label{eq:1000K_tem}
                T(r)=130\cdot (r/r_0)^{-0.7}\:\mathrm{K}.
            \end{equation}
            \begin{equation}\label{eq:1000K_h/r}
                h/r=0.05\cdot (r/r_0)^{0.15}.
            \end{equation}
    
            Regarding boundary conditions, we assume that our CPMOD model represents an isolated disk, i.e. no material from outside can cross the gap and reach the disk. 
            We set 2 ghost grids inside and outside the radial direction. The density and velocities of the ghost grids at the inner boundary are set as the interpolated values of the initial conditions \citep{Fung2014}. The velocities at ghost grids for the outer boundary are determined by $\dot{M} \equiv 2 \pi r \Sigma_\mathrm{g} v_{r} = const$, while its density uses a fixed interpolation.
            
            Regarding the dust setting of the disk, we did not separately add dust in the model. The SED we want to obtain is mainly in the infrared band of JWST/MIRI (5 to 28.5 $\mathrm{\mu m}$), and the main radiation of this magnitude comes from micron-scale dust. This part of dust grain size in the hydrodynamic model is between 0.1 to 10 micron, same as it in parametric models. Therefore, we assume that micron-scale dust and gas are well coupled, and therefore their density distribution should be constrained by the value ``dust to gas ratio'', which was set to 0.001.
            
        \begin{figure*}[t!]
            \includegraphics[width=0.95\textwidth]{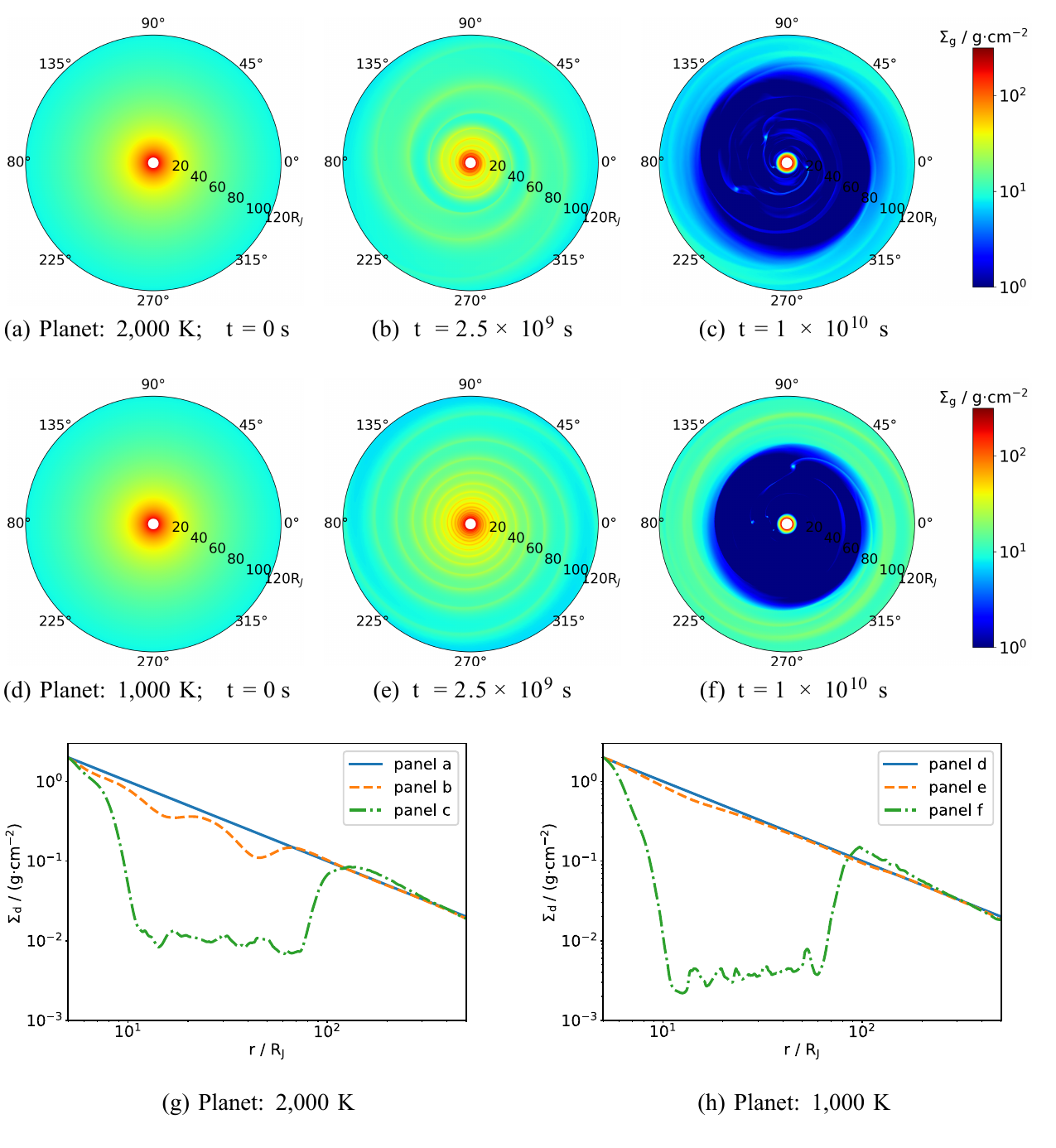}
            \caption{Snapshots of two sets of hydrodynamic simulation. The first and second rows illustrate the gas surface density distribution of 2,000 K and 1,000 K models (i.e. different isothermal temperature profiles) at three different epochs (beginning, before adding moons, and after 2,000 orbits of the outermost moon), respectively. The bottom row displays the dust density profiles as a function of radius from each panel. The dust surface density is converted from gas surface density by a constant dust-to-gas ratio.}
            \label{fig:hdresult}
        \end{figure*}
        
        \subsection{Simulations} \label{sec:hd-simulation}

            Our model is configured with 768 $\times$ 1024 grid points in the radial and azimuthal directions, respectively. The radial grid points follow a logarithmic distribution, while the azimuthal grid points are uniform.
    
            Before joining the moons, the hydrodynamic model was simulated for a time of $2.5 \times 10^9$ s. After joining the four moons, it was simulated for more than 2,000 orbits of the outermost moon (about $7.5 \times 10^9$ s) to obtain approximately stable hydrodynamic results. Then we use the density profile before moons' entry as a full disk hydrodynamic model, and the density profile after 2,000 orbits as a pre-transitional disk hydrodynamic model. We build two hydrodynamic models, and the other model just has a little difference, whose central planet is at 1,000 K, and thus the aspect ratio is different.

            The hydrodynamic simulation results of hydrodynamic models are depicted in Fig. \ref{fig:hdresult}. The first and second rows of figures contain models with the central planet at 2,000 K and 1,000 K, respectively, showing the distribution of model gas density at different evolutionary times. Since the dust particles in these models are small and well coupled to the gas, the dust distribution directly depends on the gas distribution and the dust-to-gas ratio. It can be seen that at the beginning, the models were in the continuous state of full disks (panel a and panel d). When only a star was placed on the outside of the disk and the moons were not added, the disks were subjected to tidal torque and both formed a spiral arm structure (panel b and panel e). After the moons were added, and the models were simulated for enough time, the four massive moons finally formed a gap structure in each disk (panel c and panel f).
    
            The third row of figures illustrates the evolution of the dust density profile at different times for models with the central planet at 2,000 K (panel g) and 1,000 K (panel h), respectively. The dust density profile is obtained by multiplying the gas density profile by dust-to-gas ratio 0.001. From the figures, it is evident that the stable disks with spiral arms (dashed line) maintain a density profile similar to the full disks (solid line). However, with the addition of moons, the disks eventually evolve into pre-transitional disks (dash-dotted line) exhibiting a noticeable gap. The difference of the gap structure depth in two models, are mainly caused by the different aspect ratio values, as shown in Eq.\ref{eq:2000K_h/r} and Eq.\ref{eq:1000K_h/r}, which is like the condition in PPDs\citep{Kanagawa_2015}.

\section{Constant Derivation}\label{appendix:deriv}

    The derivation of $b^\prime$ proceeds as follows:
    According to Planck's blackbody radiation formula, the radiation intensity is given by
        \begin{equation}
            I_\lambda(\lambda)=\frac{u(\lambda)}{4\pi}=\frac{2hc}{\lambda^5}\cdot\frac{1}{e^{\frac{hc}{\lambda kT}}-1}.
        \end{equation}
    Since the extremum of the radiation flux $F_\lambda$ coincides with $I_\lambda$, to derive Wien's displacement law, it is sufficient to directly differentiate $I_\lambda$ and find its extremum.
    However, what we need is the peak value of $\lambda F_\lambda$, thus differentiating $\lambda I_\lambda$, we have
        \begin{equation}
            \frac{\partial (\lambda I_\lambda (\lambda))}{\partial \lambda} = \frac{hc}{\lambda kT} \cdot \frac{e^{\frac{hc}{\lambda kT}}}{e^{\frac{hc}{\lambda kT}}-1} -4.
        \end{equation}
    Setting it equals 0, and let the dimensionless variable $x=\frac{hc}{\lambda kT}$, we get
        \begin{equation}
            \frac{xe^x}{e^x-1}-4=0.
        \end{equation}
    The analytical solution of the equation cannot be expressed in elementary functions, but an approximate solution can be obtained using numerical methods, yielding $x\approx3.92069$.
    Substituting back, we obtain the maximum point at $\lambda^\prime _\mathrm{max}\approx\: \frac{3674\mathrm{\mu m\cdot K}}{T}$, which is $b^\prime\approx\:3674\mathrm{\mu m\cdot K}$.


\bibliography{ref}{}

\begin{thebibliography}{}
\expandafter\ifx\csname natexlab\endcsname\relax\def\natexlab#1{#1}\fi
\providecommand{\url}[1]{\href{#1}{#1}}
\providecommand{\dodoi}[1]{doi:~\href{http://doi.org/#1}{\nolinkurl{#1}}}
\providecommand{\doeprint}[1]{\href{http://ascl.net/#1}{\nolinkurl{http://ascl.net/#1}}}
\providecommand{\doarXiv}[1]{\href{https://arxiv.org/abs/#1}{\nolinkurl{https://arxiv.org/abs/#1}}}

\bibitem[{Andrews {et~al.}(2021)Andrews, Elder, Zhang, Huang, Benisty,
  Kurtovic, Wilner, Zhu, Carpenter, Pérez, Teague, Isella, \&
  Ricci}]{Andrews_2021}
Andrews, S.~M., Elder, W., Zhang, S., {et~al.} 2021, The Astrophysical Journal,
  916, 51, \dodoi{10.3847/1538-4357/ac00b9}

\bibitem[{Bae {et~al.}(2022)Bae, Teague, Andrews, Benisty, Facchini,
  Galloway-Sprietsma, Loomis, Aikawa, Alarcón, Bergin, Bergner, Booth,
  Cataldi, Cleeves, Czekala, Guzmán, Huang, Ilee, Kurtovic, Law, Gal, Liu,
  Long, Ménard, Öberg, Pérez, Qi, Schwarz, Sierra, Walsh, Wilner, \&
  Zhang}]{Bae_2022}
Bae, J., Teague, R., Andrews, S.~M., {et~al.} 2022, The Astrophysical Journal
  Letters, 934, L20, \dodoi{10.3847/2041-8213/ac7fa3}

\bibitem[{Benisty {et~al.}(2021)Benisty, Bae, Facchini, Keppler, Teague,
  Isella, Kurtovic, Pérez, Sierra, Andrews, Carpenter, Czekala, Dominik,
  Henning, Menard, Pinilla, \& Zurlo}]{Benisty_2021}
Benisty, M., Bae, J., Facchini, S., {et~al.} 2021, The Astrophysical Journal
  Letters, 916, L2, \dodoi{10.3847/2041-8213/ac0f83}

\bibitem[{Betti {et~al.}(2022)Betti, Follette, Ward-Duong, Aoyama, Marleau,
  Bary, Robinson, Janson, Balmer, Chauvin, \& Palma-Bifani}]{Betti_2022}
Betti, S.~K., Follette, K.~B., Ward-Duong, K., {et~al.} 2022, The Astrophysical
  Journal Letters, 935, L18, \dodoi{10.3847/2041-8213/ac85ef}

\bibitem[{Birnstiel {et~al.}(2018)Birnstiel, Dullemond, Zhu, Andrews, Bai,
  Wilner, Carpenter, Huang, Isella, Benisty, Pérez, \& Zhang}]{Birnstiel_2018}
Birnstiel, T., Dullemond, C.~P., Zhu, Z., {et~al.} 2018, The Astrophysical
  Journal, 869, L45, \dodoi{10.3847/2041-8213/aaf743}

\bibitem[{Chen {et~al.}(2020)Chen, Yang, Martin, \& Zhu}]{Chen_2020}
Chen, C., Yang, C.-C., Martin, R.~G., \& Zhu, Z. 2020, Monthly Notices of the
  Royal Astronomical Society, 500, 2822–2830, \dodoi{10.1093/mnras/staa3427}

\bibitem[{{Chen} {et~al.}(2022){Chen}, {Bailey}, {Stone}, \&
  {Zhu}}]{2022ApJ...939L..23C}
{Chen}, Y.-X., {Bailey}, A., {Stone}, J., \& {Zhu}, Z. 2022, \apjl, 939, L23,
  \dodoi{10.3847/2041-8213/ac9b3e}

\bibitem[{{Choksi} \& {Chiang}(2024)}]{2024arXiv240310057C}
{Choksi}, N., \& {Chiang}, E. 2024, arXiv e-prints, arXiv:2403.10057,
  \dodoi{10.48550/arXiv.2403.10057}

\bibitem[{Cugno {et~al.}(2024)Cugno, Patapis, Banzatti, Meyer, Dannert,
  Stolker, MacDonald, \& Pontoppidan}]{cugno2024midinfrared}
Cugno, G., Patapis, P., Banzatti, A., {et~al.} 2024, Mid-Infrared Spectrum of
  the Disk around the Forming Companion GQ Lup B Revealed by JWST/MIRI.
\newblock \doarXiv{2404.07086}

\bibitem[{{Dullemond} {et~al.}(2012){Dullemond}, {Juhasz}, {Pohl}, {Sereshti},
  {Shetty}, {Peters}, {Commercon}, \& {Flock}}]{2012ascl.soft02015D}
{Dullemond}, C.~P., {Juhasz}, A., {Pohl}, A., {et~al.} 2012, {RADMC-3D: A
  multi-purpose radiative transfer tool}, Astrophysics Source Code Library,
  record ascl:1202.015.
\newblock \doeprint{1202.015}

\bibitem[{{Espaillat} {et~al.}(2014){Espaillat}, {Muzerolle}, {Najita},
  {Andrews}, {Zhu}, {Calvet}, {Kraus}, {Hashimoto}, {Kraus}, \&
  {D'Alessio}}]{2014prpl.conf..497E}
{Espaillat}, C., {Muzerolle}, J., {Najita}, J., {et~al.} 2014, in Protostars
  and Planets VI, ed. H.~{Beuther}, R.~S. {Klessen}, C.~P. {Dullemond}, \&
  T.~{Henning}, 497--520, \dodoi{10.2458/azu_uapress_9780816531240-ch022}

\bibitem[{{Fung} {et~al.}(2014){Fung}, {Shi}, \& {Chiang}}]{Fung2014}
{Fung}, J., {Shi}, J.-M., \& {Chiang}, E. 2014, \apj, 782, 88,
  \dodoi{10.1088/0004-637X/782/2/88}

\bibitem[{Fung {et~al.}(2019)Fung, Zhu, \& Chiang}]{Fung_2019}
Fung, J., Zhu, Z., \& Chiang, E. 2019, The Astrophysical Journal, 887, 152,
  \dodoi{10.3847/1538-4357/ab53da}

\bibitem[{Furlan {et~al.}(2006)Furlan, Hartmann, Calvet, D’Alessio,
  Franco-Hernández, Forrest, Watson, Uchida, Sargent, Green, Keller, \&
  Herter}]{Furlan_2006}
Furlan, E., Hartmann, L., Calvet, N., {et~al.} 2006, The Astrophysical Journal
  Supplement Series, 165, 568, \dodoi{10.1086/505468}

\bibitem[{Hernandez {et~al.}(2007)Hernandez, Calvet, Briceno, Hartmann, Vivas,
  Muzerolle, Downes, Allen, \& Gutermuth}]{Hernandez_2007}
Hernandez, J., Calvet, N., Briceno, C., {et~al.} 2007, The Astrophysical
  Journal, 671, 1784–1799, \dodoi{10.1086/522882}

\bibitem[{Isella {et~al.}(2019)Isella, Benisty, Teague, Bae, Keppler, Facchini,
  \& Pérez}]{Isella_2019}
Isella, A., Benisty, M., Teague, R., {et~al.} 2019, The Astrophysical Journal
  Letters, 879, L25, \dodoi{10.3847/2041-8213/ab2a12}

\bibitem[{Kanagawa {et~al.}(2015)Kanagawa, Muto, Tanaka, Tanigawa, Takeuchi,
  Tsukagoshi, \& Momose}]{Kanagawa_2015}
Kanagawa, K.~D., Muto, T., Tanaka, H., {et~al.} 2015, The Astrophysical
  Journal, 806, L15, \dodoi{10.1088/2041-8205/806/1/l15}

\bibitem[{{Lada} \& {Adams}(1992)}]{1992ApJ...393..278L}
{Lada}, C.~J., \& {Adams}, F.~C. 1992, \apj, 393, 278, \dodoi{10.1086/171505}

\bibitem[{{Liu} {et~al.}(2018){Liu}, {Jin}, {Li}, {Isella}, \&
  {Li}}]{2018ApJ...857...87L}
{Liu}, S.-F., {Jin}, S., {Li}, S., {Isella}, A., \& {Li}, H. 2018, \apj, 857,
  87, \dodoi{10.3847/1538-4357/aab718}

\bibitem[{Luhman {et~al.}(2023)Luhman, Tremblin, Birkmann, Manjavacas, Valenti,
  Alves~de Oliveira, Beck, Giardino, Lützgendorf, Rauscher, \&
  Sirianni}]{Luhman_2023}
Luhman, K.~L., Tremblin, P., Birkmann, S.~M., {et~al.} 2023, The Astrophysical
  Journal Letters, 949, L36, \dodoi{10.3847/2041-8213/acd635}

\bibitem[{{McKinnon}(2023)}]{2023ASSL..468...41M}
{McKinnon}, W.~B. 2023, in Astrophysics and Space Science Library, Vol. 468,
  Io: A New View of Jupiter's Moon, ed. R.~M.~C. {Lopes}, K.~{de Kleer}, \&
  J.~T. {Keane}, 41--93, \dodoi{10.1007/978-3-031-25670-7_3}

\bibitem[{Meru {et~al.}(2013)Meru, Galvagni, \& Olczak}]{Meru_2013}
Meru, F., Galvagni, M., \& Olczak, C. 2013, The Astrophysical Journal, 774, L4,
  \dodoi{10.1088/2041-8205/774/1/l4}

\bibitem[{Miret-Roig {et~al.}(2021)Miret-Roig, Bouy, Raymond, Tamura, Bertin,
  Barrado, Olivares, Galli, Cuillandre, Sarro, Berihuete, \&
  Huélamo}]{Miret_Roig_2021}
Miret-Roig, N., Bouy, H., Raymond, S.~N., {et~al.} 2021, Nature Astronomy, 6,
  89–97, \dodoi{10.1038/s41550-021-01513-x}

\bibitem[{{Peeters} {et~al.}(2004){Peeters}, {Spoon}, \&
  {Tielens}}]{2004ApJ...613..986P}
{Peeters}, E., {Spoon}, H.~W.~W., \& {Tielens}, A.~G.~G.~M. 2004, \apj, 613,
  986, \dodoi{10.1086/423237}

\bibitem[{{Rein} \& {Liu}(2012)}]{2012A&A...537A.128R}
{Rein}, H., \& {Liu}, S.~F. 2012, \aap, 537, A128,
  \dodoi{10.1051/0004-6361/201118085}

\bibitem[{Ringqvist {et~al.}(2023)Ringqvist, Viswanath, Aoyama, Janson,
  Marleau, \& Brandeker}]{Ringqvist_2023}
Ringqvist, S.~C., Viswanath, G., Aoyama, Y., {et~al.} 2023, Astronomy \&
  Astrophysics, 669, L12, \dodoi{10.1051/0004-6361/202245424}

\bibitem[{Santamaría-Miranda {et~al.}(2017)Santamaría-Miranda, Cáceres,
  Schreiber, Hardy, Bayo, Parsons, Gromadzki, \&
  Aguayo~Villegas}]{Santamar_a_Miranda_2017}
Santamaría-Miranda, A., Cáceres, C., Schreiber, M.~R., {et~al.} 2017, Monthly
  Notices of the Royal Astronomical Society, 475, 2994–3003,
  \dodoi{10.1093/mnras/stx3325}

\bibitem[{{Scholz} {et~al.}(2023){Scholz}, {Muzic}, {Jayawardhana},
  {Almendros-Abad}, \& {Wilson}}]{2023AJ....165..196S}
{Scholz}, A., {Muzic}, K., {Jayawardhana}, R., {Almendros-Abad}, V., \&
  {Wilson}, I. 2023, \aj, 165, 196, \dodoi{10.3847/1538-3881/acc65d}

\bibitem[{{Shakura} \& {Sunyaev}(1973)}]{1973A&A....24..337S}
{Shakura}, N.~I., \& {Sunyaev}, R.~A. 1973, \aap, 24, 337

\bibitem[{Spiegel \& Burrows(2012)}]{Spiegel_2012}
Spiegel, D.~S., \& Burrows, A. 2012, The Astrophysical Journal, 745, 174,
  \dodoi{10.1088/0004-637x/745/2/174}

\bibitem[{{Stone} {et~al.}(2020){Stone}, {Tomida}, {White}, \&
  {Felker}}]{2020ApJS..249....4S}
{Stone}, J.~M., {Tomida}, K., {White}, C.~J., \& {Felker}, K.~G. 2020, \apjs,
  249, 4, \dodoi{10.3847/1538-4365/ab929b}

\bibitem[{{Tanigawa} {et~al.}(2012){Tanigawa}, {Ohtsuki}, \&
  {Machida}}]{2012ApJ...747...47T}
{Tanigawa}, T., {Ohtsuki}, K., \& {Machida}, M.~N. 2012, \apj, 747, 47,
  \dodoi{10.1088/0004-637X/747/1/47}

\bibitem[{Walker {et~al.}(2004)Walker, Wood, Lada, Robitaille, Bjorkman, \&
  Whitney}]{Walker_2004}
Walker, C., Wood, K., Lada, C.~J., {et~al.} 2004, Monthly Notices of the Royal
  Astronomical Society, 351, 607–616,
  \dodoi{10.1111/j.1365-2966.2004.07807.x}

\bibitem[{Wang {et~al.}(2021)Wang, Vigan, Lacour, Nowak, Stolker, Rosa,
  Ginzburg, Gao, Abuter, Amorim, Asensio-Torres, Bauböck, Benisty, Berger,
  Beust, Beuzit, Blunt, Boccaletti, Bohn, Bonnefoy, Bonnet, Brandner,
  Cantalloube, Caselli, Charnay, Chauvin, Choquet, Christiaens, Clénet,
  du~Foresto, Cridland, de~Zeeuw, Dembet, Dexter, Drescher, Duvert, Eckart,
  Eisenhauer, Facchini, Gao, Garcia, Lopez, Gardner, Gendron, Genzel,
  Gillessen, Girard, Haubois, Heißel, Henning, Hinkley, Hippler, Horrobin,
  Houllé, Hubert, Jiménez-Rosales, Jocou, Kammerer, Keppler, Kervella, Meyer,
  Kreidberg, Lagrange, Lapeyrère, Bouquin, Léna, Lutz, Maire, Ménard,
  Mérand, Mollière, Monnier, Mouillet, Müller, Nasedkin, Ott, Otten,
  Paladini, Paumard, Perraut, Perrin, Pfuhl, Pueyo, Rameau, Rodet,
  Rodríguez-Coira, Rousset, Scheithauer, Shangguan, Shimizu, Stadler, Straub,
  Straubmeier, Sturm, Tacconi, van Dishoeck, Vincent, von Fellenberg,
  Ward-Duong, Widmann, Wieprecht, Wiezorrek, Woillez, \&
  Collaboration}]{Wang_2021}
Wang, J.~J., Vigan, A., Lacour, S., {et~al.} 2021, The Astronomical Journal,
  161, 148, \dodoi{10.3847/1538-3881/abdb2d}

\bibitem[{{Ward} \& {Canup}(2010)}]{2010AJ....140.1168W}
{Ward}, W.~R., \& {Canup}, R.~M. 2010, \aj, 140, 1168,
  \dodoi{10.1088/0004-6256/140/5/1168}

\bibitem[{Whitney {et~al.}(2003)Whitney, Wood, Bjorkman, \&
  Cohen}]{Whitney_2003}
Whitney, B.~A., Wood, K., Bjorkman, J.~E., \& Cohen, M. 2003, The Astrophysical
  Journal, 598, 1079–1099, \dodoi{10.1086/379068}

\bibitem[{Wolff {et~al.}(2017)Wolff, Ménard, Caceres, Lefèvre, Bonnefoy,
  Cánovas, Maret, Pinte, Schreiber, \& Plas}]{Wolff_2017}
Wolff, S.~G., Ménard, F., Caceres, C., {et~al.} 2017, The Astronomical
  Journal, 154, 26, \dodoi{10.3847/1538-3881/aa74cd}

\bibitem[{{Wright} {et~al.}(2023){Wright}, {Rieke}, {Glasse}, {Ressler},
  {Garc{\'\i}a Mar{\'\i}n}, {Aguilar}, {Alberts}, {{\'A}lvarez-M{\'a}rquez},
  {Argyriou}, {Banks}, {Baudoz}, {Boccaletti}, {Bouchet}, {Bouwman}, {Brandl},
  {Breda}, {Bright}, {Cale}, {Colina}, {Cossou}, {Coulais}, {Cracraft}, {De
  Meester}, {Dicken}, {Engesser}, {Etxaluze}, {Fox}, {Friedman}, {Fu},
  {Gasman}, {G{\'a}sp{\'a}r}, {Gastaud}, {Geers}, {Glauser}, {Gordon},
  {Greene}, {Greve}, {Grundy}, {G{\"u}del}, {Guillard}, {Haderlein},
  {Hashimoto}, {Henning}, {Hines}, {Holler}, {Detre}, {Jahromi}, {James},
  {Jones}, {Justtanont}, {Kavanagh}, {Kendrew}, {Klaassen}, {Krause},
  {Labiano}, {Lagage}, {Lambros}, {Larson}, {Law}, {Lee}, {Libralato}, {Lorenzo
  Alverez}, {Meixner}, {Morrison}, {Mueller}, {Murray}, {Mycroft}, {Myers},
  {Nayak}, {Naylor}, {Nickson}, {Noriega-Crespo}, {{\"O}stlin}, {O'Sullivan},
  {Ottens}, {Patapis}, {Penanen}, {Pietraszkiewicz}, {Ray}, {Regan},
  {Roteliuk}, {Royer}, {Samara-Ratna}, {Samuelson}, {Sargent}, {Scheithauer},
  {Schneider}, {Schreiber}, {Shaughnessy}, {Sheehan}, {Shivaei}, {Sloan},
  {Tamas}, {Teague}, {Temim}, {Tikkanen}, {Tustain}, {van Dishoeck},
  {Vandenbussche}, {Weilert}, {Whitehouse}, \& {Wolff}}]{2023PASP..135d8003W}
{Wright}, G.~S., {Rieke}, G.~H., {Glasse}, A., {et~al.} 2023, \pasp, 135,
  048003, \dodoi{10.1088/1538-3873/acbe66}

\bibitem[{Wu {et~al.}(2022)Wu, Bowler, Sheehan, Close, Eisner, Best,
  Ward-Duong, Zhu, \& Kraus}]{Wu_2022}
Wu, Y.-L., Bowler, B.~P., Sheehan, P.~D., {et~al.} 2022, The Astrophysical
  Journal Letters, 930, L3, \dodoi{10.3847/2041-8213/ac6420}

\bibitem[{{Zhang} {et~al.}(2024){Zhang}, {Huang}, \& {Dong}}]{Zhang2024}
{Zhang}, M., {Huang}, P., \& {Dong}, R. 2024, \apj, 961, 86,
  \dodoi{10.3847/1538-4357/ad055c}

\bibitem[{Zhou {et~al.}(2014)Zhou, Herczeg, Kraus, Metchev, \&
  Cruz}]{Zhou_2014}
Zhou, Y., Herczeg, G.~J., Kraus, A.~L., Metchev, S., \& Cruz, K.~L. 2014, The
  Astrophysical Journal, 783, L17, \dodoi{10.1088/2041-8205/783/1/l17}

\bibitem[{{Zhu}(2015)}]{2015ApJ...799...16Z}
{Zhu}, Z. 2015, \apj, 799, 16, \dodoi{10.1088/0004-637X/799/1/16}

\end{thebibliography}
\bibliographystyle{aasjournal}


\end{CJK*}
\end{document}